\documentclass{emulateapj}

\usepackage{rotating}

\def\gsim{\mathrel{\raise0.35ex\hbox{$\scriptstyle >$}\kern-0.6em
\lower0.40ex\hbox{{$\scriptstyle \sim$}}}}
\def\lsim{\mathrel{\raise0.35ex\hbox{$\scriptstyle <$}\kern-0.6em
\lower0.40ex\hbox{{$\scriptstyle \sim$}}}}
\def\gs{\mathrel{\raise0.35ex\hbox{$\scriptstyle >$}\kern-0.6em
\lower0.40ex\hbox{{$\scriptstyle \sim$}}}}
\def\ls{\mathrel{\raise0.35ex\hbox{$\scriptstyle <$}\kern-0.6em
\lower0.40ex\hbox{{$\scriptstyle \sim$}}}}

\slugcomment{accepted for publication in ApJ, February 28, 2010}

\shorttitle{Starburst or AGN dominance in submm-luminous candidate AGN }
\shortauthors{Coppin et al.}

\begin{document}

\title{Mid-infrared spectroscopy of candidate AGN-dominated submillimeter galaxies}

\author{K. Coppin,$^{1}$ A. Pope,$^{2,3}$ K. Men\'{e}ndez Delmestre,$^{4,5}$ D.M. Alexander,$^{6}$ J.S. Dunlop,$^{7}$ E. Egami,$^{8}$, J. Gabor,$^{8}$ Edo Ibar,$^{9}$ R.J. Ivison,$^{9,7}$ J.E. Austermann,$^{10}$ A.W. Blain,$^{11}$ S.C. Chapman,$^{12}$ D.L. Clements,$^{13}$ L. Dunne,$^{14}$ S. Dye,$^{15}$ D. Farrah,$^{16}$ D.H. Hughes,$^{17}$ A.M.J. Mortier,$^{7}$ M.J. Page,$^{18}$ M. Rowan-Robinson,$^{13}$ D. Scott,$^{19}$ C. Simpson,$^{20}$  Ian Smail,$^{1}$ A.M. Swinbank,$^{1}$ M. Vaccari,$^{21}$ M.S. Yun,$^{22}$ }

\footnotetext[]{\hspace{-0.4cm}$^{1}$ Institute for Computational Cosmology, Durham University, South Road, Durham, DH1 3LE, UK\\
$^{2}$ National Optical Astronomy Observatory, 950 N. Cherry Ave., Tucson, AZ, 85719\\
$^{3}$ \textit{Spitzer} Fellow\\
$^{4}$ NSF Astronomy and Astrophysics Postdoctoral Fellow\\
$^{5}$ The Observatories of the Carnegie Institution for Science, 813 Santa Barbara St., Pasadena, CA, 91101, USA\\
$^{6}$ Department of Physics, Durham University, South Road, Durham, DH1 3LE, UK\\
$^{7}$ Scottish Universities Physics Alliance (SUPA), Institute for Astronomy, University of Edinburgh, Royal Observatory, Blackford Hill, Edinburgh, EH9 3HJ, UK\\
$^{8}$ Steward Observatory, University of Arizona, 933 N. Cherry Ave., Tuscon, AZ, 85721\\
$^{9}$ UK ATC, Science and Technology Facilities Council, Royal Observatory, Blackford Hill, Edinburgh EH9 3HJ, UK\\
$^{10}$ Center for Astrophysics and Space Astronomy, University of Colorado, Boulder, CO 80309\\
$^{11}$ Caltech, 249-17, Pasadena, CA 91125, USA\\
$^{12}$ Institute of Astronomy, University of Cambridge, Madingley Road, Cambridge CB3 0HA, UK\\
$^{13}$ Astrophysics Group, Blackett Laboratory, Imperial College, Prince Consort Rd., London SW7 2BW, UK\\
$^{14}$ The School of Physics and Astronomy, University of Nottingham, University Park, Nottingham NG7 2RD, UK\\
$^{15}$ School of Physics and Astronomy, Cardiff University, 5, The Parade, Cardiff CF24 3YB, UK\\
$^{16}$ Astronomy Centre, University of Sussex, Falmer, Brighton BN1 9QH, UK\\
$^{17}$ Instituto Nacional de Astrof\'{\i}sica, \'{O}ptica y Electr\'{o}nica, Apartado Postal 51 y 216, 72000 Puebla, Pue., Mexico\\
$^{18}$ Mullard Space Science Laboratory (MSSL), University College London, Holmbury St. Mary, Dorking, Surrey RH5 6NT, UK\\
$^{19}$ Department of Physics \& Astronomy, University of British Columbia, 6224 Agricultural Road, Vancouver, B.C., V6T 1Z1, Canada\\
$^{20}$ Astrophysics Research Institute, Liverpool John Moores University, Twelve Quays House, Egerton Wharf, Birkenhead CH41 1LD, UK\\
$^{21}$ Department of Astronomy, University of Padova, Vicolo Osservatorio 3, I-35122, Padova, Italy\\
$^{22}$ Department of Astronomy, University of Massachusetts, Amherst, MA 01003\\
}

\begin{abstract}

\textit{Spitzer} spectroscopy has revealed that $\simeq80$\% of submm galaxies (SMGs) are starburst (SB) dominated in the mid-infrared.  Here we focus on the remaining $\simeq20$\% that show signs of harboring powerful active galactic nuclei (AGN). We have obtained \textit{Spitzer}-IRS spectroscopy of a sample of eight SMGs which are candidates for harboring powerful AGN on the basis of IRAC color-selection ($S_{8\mu\mathrm{m}}/S_{4.5\mu\mathrm{m}}>2$; i.e.~likely power-law mid-infrared SEDs).  SMGs with an AGN dominating ($\gsim50\%$) their mid-infrared emission could represent the `missing link' sources in an evolutionary sequence involving a major merger.  First of all, we detect PAH features in \textit{all} of the SMGs, indicating redshifts from 2.5--3.4, demonstrating the power of the mid-infrared to determine redshifts for these optically faint dusty galaxies.  Secondly, we see signs of both star-formation (from the PAH features) and AGN activity (from continuum emission) in our sample:  62\% of the sample are AGN-dominated in the mid-infrared with a median AGN content of 56\%, compared with $<30$\% on average for typical SMGs, revealing that our IRAC color selection has successfully singled out sources with proportionately more AGN emission than typical SB-dominated SMGs.  However, we find that only about 10\% of these AGN dominate the bolometric emission of the SMG when the results are extrapolated to longer infrared wavelengths, implying that AGN are not a significant power source to the SMG population overall, even when there is evidence in the mid-infrared for substantial AGN activity.  When existing samples of mid-infrared AGN-dominated SMGs are considered, we find that $S_{8\mu\mathrm{m}}/S_{4.5\mu\mathrm{m}}>1.65$ works well at selecting mid-infrared energetically dominant AGN in SMGs, implying a duty cycle of $\sim15$\% if all SMGs go through a subsequent mid-infrared AGN-dominated phase in the proposed evolutionary sequence.

\end{abstract}


\defcitealias{Pope08}{P08}
\defcitealias{KMD09}{MD09}

\section{Introduction}\label{sec:intro}

The era some 3\,Gyrs after the Big Bang ($z\sim 2$--2.5) coincides with a peak in activity in two important populations:  QSOs, which represent accretion onto supermassive black holes (SMBHs); and a population of extremely luminous, but highly obscured galaxies (e.g.\ \citealt{Chapman05}; \citealt{Wall08}).  The bulk of the luminosity of these obscured galaxies is emitted in the rest-frame far-infrared waveband. As a result of redshifting, they are most directly selected through their emission in the submillimeter (submm) or mm wavebands, typically in the atmospheric windows around 850 or $1100\,\mathrm{\mu m}$, and so are termed submm galaxies (SMGs).  The infrared luminosities ($L_\mathrm{IR}$) inferred for SMGs from their submm emission are highly uncertain, but assuming, as appears to be the case, that they follow the far-infrared--radio correlation for local starburst (SB) galaxies, then their typical luminosities will be $L_\mathrm{IR}\simeq10^{12}$--$10^{13}\,\mathrm{L}_\odot$ (e.g.~\citealt{Kovacs06}; \citealt{Murphy09}).  Thus this population may contain some of the most luminous galaxies in the Universe comparable in luminosity to QSOs.   

The increasing availability of precise redshifts for samples of SMGs (e.g.~\citealt{Chapman05}; \citealt{Eales09}) has allowed their properties to be studied in detail.  SMGs are  strongly clustered \citep{Blain04}, massive ($M_\star > 10^{11}$\,M$_\odot$, \citealt{Borys05}; \citealt{Hainline09}), and gas rich ($f_{gas}\sim 0.3$, \citealt{Frayer98}; \citealt{Greve05}; \citealt{Tacconi06,Tacconi08}) systems which are known to harbor (apparently) low-luminosity Compton-thin AGN (i.e.~$N_\mathrm{H}<10^{24}$\,cm$^{-2}$,  \citealt{Takata06}; \citealt{Alexander05nat,Alexander05,Alexander08a}) and (apparently) strong star-formation (SF) activity ($\mathrm{SFR}\sim 1000$\,M$_\odot$\,yr$^{-1}$; \citealt{Swinbank04}; \citealt{Chapman05}).  Many of these properties, and the similarity between the redshift distributions of QSOs and SMGs \citep{Chapman05}, support a link between SMGs, QSOs and the formation phase of massive elliptical galaxies (e.g.~\citealt{Lilly99}; \citealt{Archibald02}; \citealt{Stevens05}; \citealt{Coppin08b}).  In the high-redshift interpretation of the evolutionary sequence first presented by \citet{Sanders88}, SMGs would trace an infrared ultraluminous phase followed by a short `transition phase' where the galaxy would display a mix of SF and obscured AGN activity before evolving into an optically luminous QSO.  Studying the relative SF and AGN activity in these `transition' or `missing link' sources and performing a comparison to typical SF SMGs can thus provide important insight on the validity of this evolutionary sequence.  Here we focus on the comparison of the mid-infrared spectral properties between existing samples of typical SMGs and a new sample of candidate `transition phase' SMGs.  The benefit of undertaking this energy audit in the mid-infrared is that the spectra can provide measurements of both SF and AGN activity simultaneously (see \citealt{Pope08} and \citealt{KMD09}; hereafter \citetalias{Pope08} and \citetalias{KMD09}), providing a complementary approach to radio (e.g.~\citealt{Ibar09b}) and X-ray studies of SMGs (e.g.~\citealt{Alexander05}; \citealt{Laird09}).  

The {\it Spitzer} InfraRed Spectrograph (IRS; \citealt{Houck04}) era enabled the study of the energetics of $\approx45$ SMGs with $24\,\mathrm{\mu m}$ flux densities as low as $\sim0.1$\,mJy and with redshifts as high as $z=2.6$, demonstrating that accurate redshifts for dust-enshrouded (and sometimes optically invisible) galaxies can be obtained (e.g.~\citealt{Lutz05}; \citealt{Valiante07}; \citealt{MD07}; \citetalias{KMD09}; \citetalias{Pope08}).  The majority of the IRS work has confirmed that SMGs are primarily SB-dominated systems, with hot dust continuum from an AGN contributing at most 30\% of the mid-infrared luminosity.  Only about 15\% of blank-field SMGs appear to be continuum-dominated in the mid-infrared (i.e.~$>50$\% contribution; \citetalias{Pope08}), with InfraRed Array Camera (IRAC; \citealt{Fazio04}) 8.0\,$\mu$m to 4.5\,$\mu$m color ratios of $S_{8}/S_{4.5}>2$.  Mid-infrared continuum-dominated SMGs are potentially an important sub-population of SMGs representing the `missing link' sources in the proposed evolutionary sequence of \citet{Sanders88}, but being the minority of this luminous population, have not yet been studied in a systematic or statistically robust way. Thus, an important question to address is: \textit{Are the energetics of these composite SB/AGN objects with submm emission dominated by SBs or by AGN, even when the IRAC colors indicate that an AGN is likely present?}

\begin{figure*}
\epsscale{.80}
\plotone{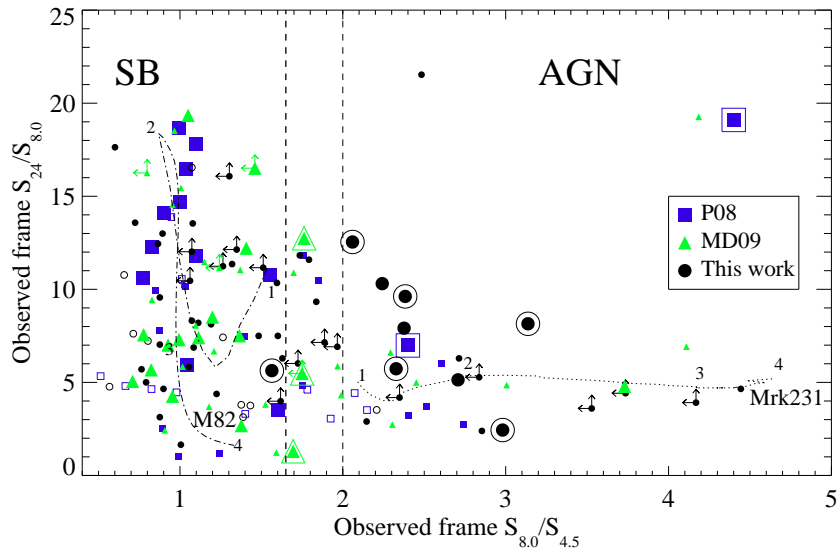}
\caption{{\it Spitzer} color-color diagram as an AGN diagnostic (see e.g.~\citealt{Ivison04}; \citetalias{Pope08}). The dotted and dot-dashed curves show the positions of Mrk231 (an AGN) and M82 (a SB), respectively, as a function of redshift (redshift is indicated by the numbers along the tracks).  The smallest filled and open circles represent robust ($P<0.05$) and tentative ($P>0.05$) 24\,$\mu$m SMG counterparts, respectively, for our parent sample of blank-field SHADES SMGs \citep{Ivison07}.  For comparison, we have also plotted the GOODS-N SMG sample (blue squares; \citealt{Pope06}), and also the \citet{Chapman05} radio-selected spectroscopically-confirmed sample of SMGs (green triangles).  SMGs from these parent samples that have been observed with the IRS (\citetalias{Pope08}; \citetalias{KMD09}; this work) are indicated by large versions of the corresponding symbols. Of these IRS-observed SMGs, those with $>50\%$ contribution from continuum (AGN) emission to the mid-infrared luminosity are circumscribed and tend to lie far from the M82 SB sequence.  Although our sample was initially selected on the basis of $S_{8}/S_{4.5}>2$ (thin dashed vertical line), when all the results from the literature are taken into account a selection of $S_{8}/S_{4.5}\gtrsim1.65$ thus seems to pick out AGN fairly efficiently, with some small amount of scatter over this boundary (thick dashed vertical line; see \S~\ref{sec:discuss}).  Extrapolating from this result implies that $\approx15$\% of blank-field SMGs will likely have an enhanced dominant AGN component in the mid-infrared.\label{fig:selection}}
\end{figure*}

Other relevant IRS samples include near-infrared-selected SBs, X-ray-selected AGN \citep{Weedman06}, and \textit{Spitzer}-selected $0.3\lsim z \lsim 3$ ULIRGs (\citealt{Yan07}; \citealt{Sajina07}; \citealt{Farrah08}; \citealt{Dasyra09}; \citealt{HC09}).  AGN-dominated sources in these samples tend to reside in a distinct parameter space in \textit{Spitzer} color-color diagrams, which is consistent with color-redshift evolution tracks of well-known local AGN (see e.g.~Fig.~\ref{fig:selection} ; \citealt{Ivison04}; \citealt{Ashby06}; \citealt{Hainline09}).   The reason why $S_{8}/S_{4.5}>2$ should locate mid-infrared AGN-dominated sources is simple:  seeing an enhanced 8\,$\mu$m flux density compared with 4.5\,$\mu$m is expected if a source has significant thermal power-law emission from an AGN accretion disk, which can dilute both the PAHs and the H-opacity minimum (1.6\,$\mu$m stellar bump, which these channels trace). Although some contamination is expected: SF-dominated sources can also show enhanced $S_{8}/S_{4.5}$ at $z\gtrsim4$ (when these channels begin to sample over the peak of stellar photospheric emission) aswell as at $z\lesssim1$ (since these channels are not yet climbing up the restframe 1.6\,$\mu$m stellar bump).  The color cut is thus appropriate for separating mid-infrared SB- and AGN-dominated SMGs from $z\simeq1$--4, which is well matched to the known redshift distribution of the SMG population.  

We have thus selected eight SMGs from the Submillimeter Common-User Bolometer Array (SCUBA; \citealt{Holland99}) HAlf Degree Extragalactic Survey (SHADES; \citealt{Mortier05}; \citealt{Coppin06}; \citealt{Austermann09}) within this relatively unexplored $S_{8}/S_{4.5}>2$ parameter space that are likely harboring AGN.  These were targeted with the IRS in order to determine the relative contribution of power-law/AGN emission versus PAH/SF emission to their power output, enabling a comparison to similar IRS samples of more typical SF-dominated SMGs.  We use these data to obtain independent redshift estimates as well as to test if the IRAC-color criterion is a secure means of pre-selecting SMG counterparts with an enhanced AGN component compared to typical SMGs -- the `missing link' sources in the proposed evolutionary sequence we wish to investigate.  

This paper is organized as follows. The sample selection, {\it Spitzer}-IRS observations, data reduction, and analysis approach are described in \S~\ref{sec:obsdr}. In \S~\ref{sec:results} we present the main results of the IRS spectroscopy, including redshifts, spectral decomposition and AGN classification, and full SED fits to determine their total infrared luminosities.  We discuss the implications that our results and those of other IRS SMG studies have on the role that SMGs play in galaxy evolution in the framework of the proposed evolutionary sequence in \S~\ref{sec:discuss}. Finally, our conclusions are given in \S~\ref{sec:concl}.

Here, we discuss the observed `mid-infrared' spectral properties of our SMGs probed by the IRS which, at our source redshifts of $z>2.5$, traces $\sim4$--10\,$\mu$m in the rest-frame.  All magnitudes in this paper are on the AB system, unless otherwise stated. We adopt cosmological parameters from the \textit{WMAP} fits \citep{Spergel03}: $\Omega_\Lambda=0.73$, $\Omega_\mathrm{m}=0.27$, and $H_\mathrm{0}=71$\,km\,s$^{-1}$\,Mpc$^{-1}$.


\section{Sample Selection, Observations, Data Reduction, and Analysis}\label{sec:obsdr}

\subsection{Sample Selection and Multi-wavelength Properties}\label{sec:sample}

SHADES provides a suitably large parent sample of SMGs from which we can select a representative and large enough sample of SMG AGN-candidates. SCUBA surveyed $\sim 0.25\,\mathrm{deg^{2}}$ at $850\,\mathrm{\mu m}$ to an rms of $\sim2$\,mJy, uncovering 120 SMGs $>\!3.5\,\sigma$ in the Lockman Hole East (LH) and the equatorial Subaru-{\it XMM} Deep Field (SXDF), centred at J2000 RA=\(2^{\mathrm{h}}17^{\mathrm{m}}57{\fs}5\), Dec=\(-5^{\circ}00'18{\farcs}5\) and RA=\(10^{\mathrm{h}}52^{\mathrm{m}}26{\fs}7\), Dec=\(57^{\circ}24'12{\farcs}6\), respectively \citep{Coppin06}.  The SHADES fields were subsequently mapped to their full $\sim 0.5\,\mathrm{deg^{2}}$ coverage by the Astronomical Thermal Emission Camera (AzTEC; \citealt{Wilson08}) at 1100\,$\mu$m to an rms depth of $\sim1$--1.5\,mJy, yielding 114 SMGs above $3.6\,\sigma$ \citep{Austermann09}, a population akin to the well-known submillimeter or SCUBA galaxies (e.g.~\citealt{Chapin09}).  A detailed comparison of the overlap in the SCUBA and AzTEC SHADES catalogs will be presented in Negrello et al.\ (in preparation).

\citet{Ivison07}, \citet{Ibar09,Ibar09b} and Arumugam et al.\ (in preparation) have identified 1.4\,GHz Very Large Array (VLA), 610\,MHz Giant Metre-wave Radio Telescope (GMRT), and/or Mid-Infrared Photometer for \textit{Spitzer} (MIPS; \citealt{Rieke04}) 24\,$\mu$m counterparts for 65\% of the SHADES SMGs.  Here we select SMGs with statistically robust 24\,$\mu$m counterparts with $S_{24\,\mathrm{\mu m}}>0.2$\,mJy (ensuring that we achieve adequate signal-to-noise, S/N, on the continuum in the IRS spectra) and observed IRAC colors of $S_{8}/S_{4.5}>2$ (\citealt{Dye08}; \citealt{Clements08}; see Fig.~\ref{fig:selection}).  Our final sample is comprised of eight sources: four SMGs from each of the SHADES SCUBA and AzTEC surveys.

Our selection criteria apply to approximately 15--25\%\footnote[1]{The range in this fraction merely represents the fact that several of the SMGs are undetected 8\,$\mu$m and so they could lie on either side of the IRAC color-cut boundary in Fig.~\ref{fig:selection}.} of the SHADES SMGs (or equivalently to 7--13\% of SMGs with 24\,$\mu$m counterparts).  Note that including the `non-robust' $24\,\mathrm{\mu m}$ IDs in these estimates gives similar fractions.  In addition, deeper radio and 24\,$\mu$m data would likely yield identifications for all of the SMGs; and there is no evidence in support of SMGs fainter than 200$\,\mu$Jy at 24$\,\mu$m being fundamentally different than perhaps lying at slightly higher redshifts (e.g.~\citealt{Pope06}).  Thus, the IRAC color-selected subset of SMGs studied here should be representative of $\sim15$\% of the blank-field SMG population.  We discuss the possibility of more subtle selection effects in \S~\ref{sec:bias}.

We also observed a 24\,$\mu$m counterpart (LOCK850.41-2) of an interesting SMG with multiple robust radio and MIPS counterparts, where the other MIPS counterpart (LOCK850.41-1) was previously observed by \citetalias{KMD09}.  We present these data separately in Appendix~\ref{app:notes}, since this source does not make the original color-cut criterion for our sample.

We make use of additional multi-wavelength data-sets available for the 24\,$\mu$m SHADES SMG counterparts in order to further interpret our results in the wider context of the entire SMG population and other high-redshift galaxy populations. Available optical--to--radio photometry for each source is given in Table~\ref{tab:targets} and was retrieved from \citet{Dye08}, \citet{Furusawa08}, and Mortier et al.\ (in preparation; Subaru $BRi'z$), \citet{Lawrence07} and Warren et al.\ (in preparation; $J$ and $K$ from the DR3 UKIRT Infrared Deep Sky Survey release -- UKIDSS, Egami et al.\ (priv. comm.), and Dunlop et al.\ (in preparation; 3.6--8\,$\mu$m \textit{Spitzer}-IRAC).

The SHADES fields also possess moderately sensitive \textit{XMM-Newton} imaging with exposures of 673\,ks in the LH \citep{Brunner08}, and 18--83\,ks in the SXDF \citep{Ueda08}.  In the deeper X-ray data, even X-ray non-detections, when combined with IRS data, can be sufficient to imply 1--2 orders of magnitude of obscuration at rest-frame 2--10\,keV and can identify Compton-thick AGN. Given the combined positional uncertainties of the 24\,$\mu$m and X-ray sources (yielding a combined positional uncertainty of 3--4$''$), we search for X-ray counterparts to the SMGs using a 4$''$ search radius.  For any X-ray detections, we extract a 0.5--2\,keV flux and calculate the rest-frame 2--10\,keV X-ray luminosity, assuming a spectral slope of $\Gamma=1.4$ (the average X-ray spectral slope for sources of this approximate X-ray flux; see Fig.~8 of \citealt{Alexander01}) to make a small K-correction ($\lsim30$\%). For the X-ray non-detections, we have derived 3\,$\sigma$ upper limits.   The X-ray fluxes and flux upper limits are given in Table~\ref{tab:agn}. 

It happens that spectroscopic redshifts have not been obtained for the majority of the SMG sample, undoubtedly due to the faintness of the corresponding optical counterparts (see Table~\ref{tab:targets}).  All of our SMGs are $R_\mathrm{AB}\gtrsim24$ (except for AzLOCK.10 which is $R_\mathrm{AB}=22.65$, and which has not yet been attempted in a spectroscopic follow-up program, to our knowledge), for which \citet{Chapman05} find an increased spectroscopic redshift failure rate.  Therefore it is not a surprise that only one of our targets has a published tentative optical spectroscopic redshift available for comparison with the IRS-derived redshift (see Appendix~\ref{app:notes}).

\subsection{\textit{Spitzer}-IRS Observations}\label{sec:obs}
The \textit{Spitzer}-IRS observations (PID 50183) were taken in spectral mapping mode, with the target being placed at six positions (separated by $20''$) along the slit using multiple cycles of the longest ramp setting of 120\,s to maximize S/N in unit time, while ensuring internal robustness against cosmic rays and the identification of rogue pixels (see \citealt{Teplitz07}).  We observed the SMGs using Long-Low 1 (LL1; 19.5--38.0\,$\mu$m) and Long-Low 2 (LL2; 14.0--21.3\,$\mu$m) to ensure that the spectra cover a significant fraction of the full range of PAHs (from 6--17\,$\mu$m in the rest-frame).  Integration times were calculated on a source-by-source basis using the observed 24\,$\mu$m fluxes to yield a S/N of $>3$--4 in the continuum near the band centres (i.e.~similar-quality spectra to those in \citetalias{Pope08} and \citealt{MD07,KMD09}).  We used high-accuracy blue IRS peak-up acquisition on nearby isolated moderately bright Two Micron All Sky Survey stars (2MASS; \citealt{Skrutskie06}).  To verify the  calibration in the LL2 spectra we also obtained $16\,\mathrm{\mu m}$ IRS peak-up imaging with $2\times30$\,s cycles in a 5-point dither for each target for an additional 1.7\,hrs in total.  Table~\ref{tab:obs} gives our target list and integration times for the spectroscopy.  Our observations were obtained in 2008 May, June (LH targets), and September (SXDF targets) in a total of 58.4\,hrs.

\subsection{Data Reduction}\label{sec:dr}

We begin with the \textit{Spitzer} pipeline (Version S18) basic calibrated data. In reducing the IRS data, we follow the same approach used to reduce IRS spectra of faint sources outlined in \citetalias{Pope08}. To summarize, we identify and clean rogue pixels in the 2D files, fit and subtract latent build-up on the arrays, perform sky subtraction by creating a normalized `supersky' for each AOR in which all other sources have been masked, and co-add all the 2D sky-subtracted data files for each nod position. We extract the 1D spectrum from the 2D co-added files for each nod using the the optimal-extraction mode in SPICE\footnote{Available at http://ssc.spitzer.caltech.edu/postbcd/spice.html}. 
Along with the target spectrum we also extract a residual sky spectrum from each 2D co-add file,  which we use to determine uncertainties on our final 1D spectra. We trim all the spectra to $<35.0$\,$\mu$m in order to restrict the analysis to the reliable and least noisy portion of the data. The observed spectra are shown in Fig.~\ref{fig:alex1}.

\begin{figure}
\epsscale{1.0}
\plotone{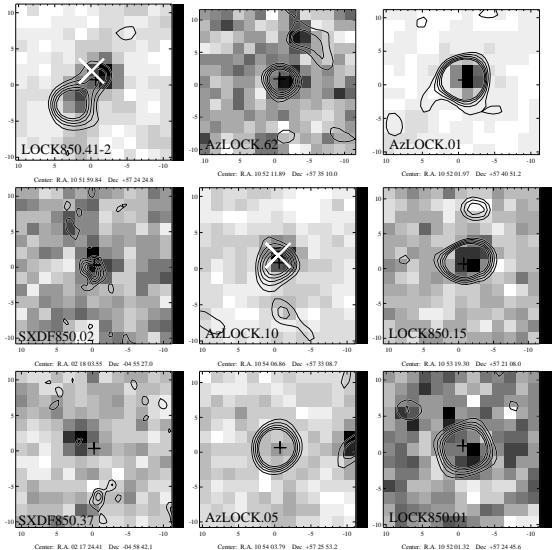}
\caption{New 16\,$\mu$m $10^{\prime\prime}\times10^{\prime\prime}$ imaging (roughly the size of the aperture of the IRS observations) centered on the 24\,$\mu$m positions of the nine SMG targets (in order of increasing redshift; see Table~\ref{tab:agn}; left-right/top-bottom), with VLA 2\,$\sigma$,3\,$\sigma$,... contours overlaid.  Source names are given in each panel (see Table~\ref{tab:obs}).  The sample has been selected on the basis of being 24\,$\mu$m detected and IRAC color-tuned in order to pick out potential AGN-dominated SMGs, although consequently the majority of the targets also appear to be radio-bright, which is consistent with an AGN-biased sample.  \textit{XMM-Newton} detections are shown by the purple `X' symbols, and only exist for the minority of the sample -- deeper X-ray data would likely yield a higher detection fraction, unless the sources are Compton-thick (see text).\label{fig:stamps}}
\end{figure}

We process the IRS 16\,$\mu$m peak-up images (PUIs) with the SSC pipeline S18. We then correct for latent charge accumulation in the PUIs by subtracting the mode flux of the central 30$\times$45 pixels from each individual exposure. We eliminate residual sky background by subtracting from each science exposure a median sky, created from median-collapsing all charge-corrected exposures and scaled to match off-source background level for each exposure. All sky-subtracted, charge-corrected science exposures are mosaicked into a single image using the standard MOPEX mosaicking pipeline, keeping the native PUI resolution scale of $1.8''$/pix. We derive 16\,$\mu$m photometry by performing point-response function (PRF) fitting analysis with the MOPEX/APEX single-frame pipeline, relying on the blue PUI PRF\footnote{Available at http://ssc.spitzer.caltech.edu/irs/puipsf}. We use `peak' thresholding as part of the image segmentation algorithm, setting the detection threshold to a 2$\sigma$ level within a 5$\times$5 pixel fitting area to ensure adequate detection of faint sources. To characterize the uncertainties in the derived integrated fluxes we rely on the integrated noise within an on-source $5\times5$ pixel aperture in the uncertainty mosaic. The 16\,$\mu$m imaging photometry is reported in Table~\ref{tab:targets} and shown in Fig.~\ref{fig:stamps}. 

We verify the calibration of our final 1D spectra by comparing to the MIPS 24\,$\mu$m and IRS/PUI 16\,$\mu$m fluxes. In all cases the spectra are consistent with the imaging photometry within the uncertainties.

\subsection{Analysis}\label{sec:analysis}

Our aim is to determine if our mid-infrared spectra show signs of AGN-dominance, and if so, to then compare these spectral properties with the overall SMG population.  We have thus followed a similar analysis approach to \citetalias{Pope08} in order to facilitate a direct unbiased comparison with existing \textit{Spitzer}-IRS samples of SMGs.  We have organised the paper so that the following subsections contain all the salient details of our analysis approach, with the results presented separately in \S~\ref{sec:results}.   In summary, in \S~\ref{sec:analysis1}, we identify the PAH features present in our spectra and determine redshifts.  We then decompose each individual spectrum into SF and AGN components in \S~\ref{sec:analysis2} in order to classify the fractional contribution of AGN to the mid-infrared luminosity. In \S~\ref{sec:analysis3} we extrapolate our results to determine the full mid--to--far-infrared SEDs in order to estimate the AGN contribution to the bolometric luminosity of these systems.

%
%
%
%
%

\begin{figure}
\epsscale{1.0}
\plotone{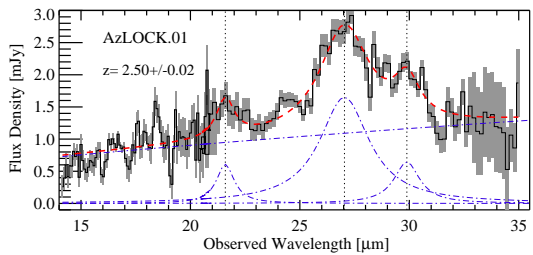}
\caption{An example of the spectral line decomposition of an SMG IRS spectrum, used to compute the redshift for the source and to measure the line profile features (see \S~\ref{sec:analysis1}). The solid black curve is the raw (unsmoothed) IRS spectrum, while the shaded region represents the associated $1\,\sigma$ noise from the sky background. The red dashed curve shows the best-fit SED to the mid-infrared spectrum, comprising a strong continuum component (blue dot-dashed power-law) and PAH features (blue dot-dashed Lorentzian profiles) at 6.62, 7.71, and 8.61\,$\mu$m (indicated by vertical lines).\label{fig:spec}}
\end{figure}
%
%
%
%
%
%
\begin{figure*}
\epsscale{1.0}
\plotone{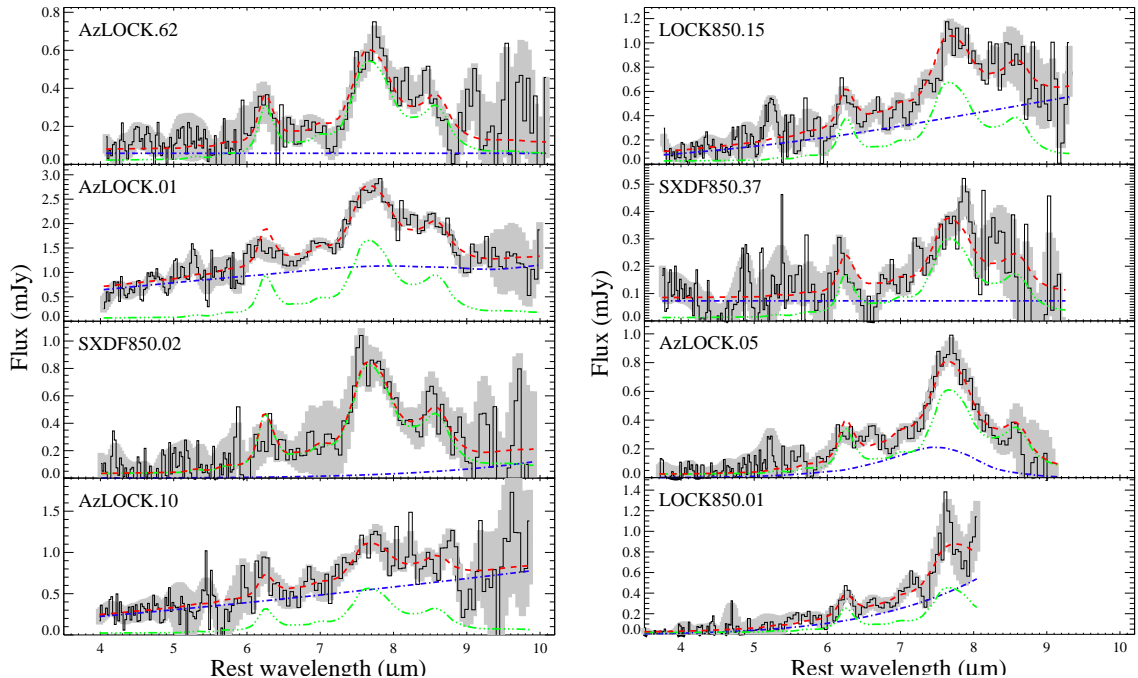}
\caption{IRS spectra of our AGN-candidate sample of SMGs, listed in order of increasing redshift and shifted to rest-frame wavelengths in order to facilitate an easy inter-comparison of the spectral features. Source names are given in each panel (see Table~\ref{tab:obs}).  The solid black curves are the raw (unsmoothed) IRS spectra, while the shaded region represents the associated $1\sigma$ noise from the sky background (see \S~\ref{sec:dr}).  The red dashed line is the best-fitting model (see \S~\ref{sec:analysis2}), which is composed of an M82 PAH template (green dot-dot-dot-dashed line) and a power-law component (blue dot-dashed line) with extinction applied.  The continuum (AGN) fraction for each source is calculated from the spectral decomposition (see Table~\ref{tab:agn}).  The spectra and fitting reveal that our sample of candidate AGN-dominated SMGs all show PAH features atop a range of mid-infrared continuum slopes, indicating a mix of SB and AGN properties in the sample.\label{fig:alex1}}
\end{figure*}

\subsubsection{Redshift Determination and Spectral Line Measurements}\label{sec:analysis1}

In order to derive the source redshifts, we simultaneously fit a power law continuum component and Lorentzian profiles to the spectra using a Levenberg-Marquardt least-squares method, including the residual sky errors in the fit determination (e.g.~\citealt{Sajina07}), assuming the following rest-frame wavelengths for the main PAH features that are visible: 6.22, 7.71, and 8.61$\,\mu$m \citep{Draine07}. We do not impose any joint constraints on the line ratios or centres (i.e.~to all lie at the same redshift), as we aim to obtain individual redshift estimates from each PAH feature. See Fig.~\ref{fig:spec} for an example of these fitting results. Our fitting routine returns the formal 1\,$\sigma$ statistical error estimates for each set of best-fitting parameters derived from the square root of the diagonal elements of the covariance matrix, and we adopt these errors in the ensuing analysis.  We verify that these formal errors are reasonable through Monte Carlo simulations similar to those described in \citetalias{Pope08}.

The final redshift for each SMG is calculated by taking a weighted average of the individual redshift estimates from the visible PAH lines (although we neglect the 8.61\,$\mu$m PAH feature in the redshift estimate because it appears relatively weak and significantly blended with the much stronger neighboring 7.71\,$\mu$m feature).  The total error on the redshift is calculated from 1\,$\sigma=(\sum_{i=1}^{n} \sigma_{PAH_i}^{-2})^{-1/2}$ with an additional `centroiding error' (calculated as the maximum spread in the line centering from the Monte Carlo simulations) added in quadrature.  For sources with only one PAH line, the 1\,$\sigma$ redshift error is taken to be the centroiding error derived from the simulations (the more conservative choice).   Note that the redshift errors range from 0.01--0.05, consistent with the similar-quality spectra in \citetalias{Pope08} and \citetalias{KMD09}.  For comparison, we use the \citet{Brandl06} SB template with fixed PAH linewidths and positions and find that this gives redshift estimates consistent with our method (which is essentially equivalent to performing a combined fit of all the lines simultaneously, but with a more accurate measure of the uncertainty of the redshift).

Once redshifts have been obtained (see Table~\ref{tab:agn}), we calculate the rest-frame PAH line equivalent widths (EWs) and luminosities from our Lorentzian model fits, propagating through the redshift and parameter errors from the fits (see Table~\ref{tab:lines}).  

\subsubsection{Determining the Relative AGN and SF Contributions in the Mid-Infrared}\label{sec:analysis2}

We determine the relative AGN and SF contributions by fitting each spectrum with a model comprised of three main components (extinction, PAHs and a continuum power-law) of the form $F_{\nu}=c_{0}\,\nu^{-c_{1}}\,e^{-c_{3}\tau_{\nu}}+c_{2}\,f_{\nu,\mathrm{M82}}$ using a $\chi^2$ minimization approach (see \citetalias{Pope08})\footnote{In contrast to the more typical SMGs from \citetalias{Pope08}, we found that the fitting never required additional extinction of the PAH component, so this term has been omitted from our model (cf.~eqn.~1 in \citetalias{Pope08}).}.  The AGN emission is characterized by a power-law with both the normalization and slope as free parameters, and $\tau_{9.7}$ is obtained from the \citet{Draine03} extinction curves.  We assume that all of the PAH line emission is powered by a SB (e.g.~\citealt{Rigopoulou99}; \citealt{Laurent00}), and the SB/PAH emission is fitted using the mid-infrared spectrum of M82 from \citet{Forster03}, which is known to fit and describe the mid-infrared spectra of the SB-dominated SMG population well (e.g.~\citetalias{Pope08}).  We derive the fraction of the mid-infrared luminosity that comes from the continuum component by integrating the continuum portion of the best-fitting model and dividing by the area under the total fit.  The use of different templates was explored by \citet{Sajina07} and \citetalias{Pope08} (e.g.~using NGC7714 in place of M82), and these authors find that it makes little difference on the final outcome of the decomposition, especially given the small wavelength range covered by the spectra.   In addition, we perform the decomposition using the \citet{Brandl06} SB template in place of M82 and find that they yield comparable fractions.

\subsubsection{Full SED Fitting and Determination of $L_\mathrm{IR}$}\label{sec:analysis3}

One of the main goals is to determine if our sample of SMG mid--to--far-infrared SEDs resemble that of typical SMGs or if an additional AGN component is required to fit the data.  With this goal in mind, using the same simple templates and approach as \citetalias{Pope08} will enable a direct and fair comparison between the two samples.  We have thus used the full mid-infrared--to--radio composite SED from \citetalias{Pope08}, comprised of their mid-infrared SMG SB composite spectrum spliced with the best-fitting modified \citet{Chary01} template (CE01 hereafter), to represent the SF component, and we have used the SED of Mrk\,231, consisting of its IRS spectrum \citep{Armus07} spliced together with the CE01 template that best fits the far-infrared/submm photometry of Mrk\,231, to represent the AGN component.  Although the resulting decomposition will thus depend heavily on the templates we have assumed, these choices will allow us to determine the \textit{relative} differences between the spectral decomposition of our sample of AGN-dominated SMGs and of more typical SB-dominated SMGs from \citetalias{Pope08}.  Nevertheless, we discuss and explore other choices for the SB and AGN templates \S~\ref{sec:fits} to verify the dependence of our results on the assumed templates.

We fit linear combinations of these templates to the data, including the mid-infrared IRS spectra, the 850 or 1100\,$\mu$m photometry, and, where available, 350\,$\mu$m photometry from \citet{Coppin08a} using a Levenberg-Marquardt least-squares approach.  We also verify that available 70 and 450\,$\mu$m 3\,$\sigma$ upper limits from \citet{Hainline09} and \citet{Coppin06}, respectively, are not violated in the final fits.  We then integrate the total best-fitting linear combination of the SB and AGN SED components in the rest-frame from 8--1000$\,\mu$m to obtain the total infrared luminosities, $L_\mathrm{IR}$, and calculate the relative fractions and luminosities of the SB ($L_\mathrm{SB}$) and AGN ($L_\mathrm{AGN}$) components to the total infrared luminosity (see Table~\ref{tab:lines}).  Note that the full SF-SMG composite SED from \citetalias{Pope08} already contains a 30\% AGN contribution in the mid-infrared (from 5--11.5\,$\mu$m in the rest-frame), which is taken into account in our calculations of $L_\mathrm{AGN}$ and $L_\mathrm{SB}$.

After the level of AGN contamination in each SMG has been determined in this way, we calculate the SFRs for our objects from $L_\mathrm{SB}$ following \citet{Kennicutt98}, assuming a starburst less than 100\,Myr in age with a \citet{Salpeter55} initial mass function.

%
%
%
%
%
%

\section{Results}\label{sec:results}
We now present the results of our experiment, including the source spectra and redshifts (\S~\ref{sec:redshifts}), the spectral decomposition and a comparison with the X-ray view in order to gather a complete census of the energetics of the sample (\S~\ref{sec:relatives}), the construction of a composite AGN-dominated SMG (\S~\ref{sec:decomp}), and finally the full SED fits (\S~\ref{sec:fits}).  We stress that we have followed similar recipes to those of \citetalias{Pope08} in order to facilitate a direct comparison with samples of more typical SF-dominated SMGs (see \S~\ref{sec:analysis}).

\subsection{Source Spectra and Redshifts}\label{sec:redshifts}

We present the final reduced spectra and fits in Fig.~\ref{fig:alex1}.  At a glance, the spectra reveal that our sample of SMGs possess PAH features atop a range of mid-infrared continuum slopes.
Despite the strong continuum in many of the spectra, measurable PAH features are apparent in \textit{all} of our spectra, and we use these to extract redshifts (see \S~\ref{sec:analysis1} and Table~\ref{tab:agn}). Near-infrared spectroscopic redshifts have recently been obtained for SXDF850.02 and SXDF850.37 (Alaghband-Zadeh et al.\ in prep.) of $z=2.518\pm0.001$ (showing narrow lines) and $z=2.769\pm0.001$ (showing broad $H\alpha$ and O{\sc iii}5007), respectively, in agreement with the IRS-derived redshifts.  Finally, we note the suggestion by \citet{Ivison05} that a galaxy at $z_{\rm spec}=2.15$, 3\,$''$ west of LOCK850.1 may be associated with the SMG.  Our \textit{Spitzer}-IRS result suggests otherwise, and we discuss this source further in Appendix~\ref{app:notes}.  

The redshift distribution of our color-selected sample of eight objects ranges from $z=2.5$--3.4, with a median of 2.76.  This is noticeably shifted to higher redshifts compared to the \citetalias{Pope08} and \citetalias{KMD09} samples, which have a median redshift of 2.0, which is more consistent with the known redshift distribution for radio-identified SMGs with $S_{850}>5$\,mJy and $z\sim2.2$ \citep{Chapman05}.  In \S~\ref{sec:bias}, we suggest that the cause for this overall redshift difference between the sub-samples is likely due to selection effects.  

\subsection{Spectral Decomposition of Individual Galaxies}\label{sec:relatives}

\subsubsection{Relative AGN/SF Contribution in the Mid-Infrared}\label{sec:rel}

Using the fitting procedure described in \S~\ref{sec:analysis2} to determine the relative AGN/SF contribution in the mid-infrared, we have tabulated the continuum (AGN) fraction for each SMG in Table~\ref{tab:agn}.  We find a wide range of continuum fractions in our sample -- from sub-dominant to significant -- with a median continuum contribution to the mid-infrared of $\approx56$\%.  SXDF850.02 and SXDF850.37 have recently been spectroscopically confirmed at restframe optical wavelengths by Alaghband-Zadeh et al.\ (in prep.), and the restframe optical line widths are in broad agreement with our mid-infrared AGN fractions of 10\% (where narrow lines suggest negligible AGN contribution) and 62\% (where broad $H\alpha$ and O{\sc iii}5007 lines suggest the presence of an AGN), respectively.   $5/8$ (62\%) SMGs in our sample are clearly continuum-dominated sources in the mid-infrared ($\gtrsim50\%$ of the mid-infrared luminosity).  For comparison, only 2/13 (15\%) SMGs from \citetalias{Pope08} have $>50$\% AGN contribution in the mid-infrared, and the median AGN contribution of their sample (including those with upper limits) is $\approx34$\%.  

\begin{figure}
\epsscale{1.0}
\plotone{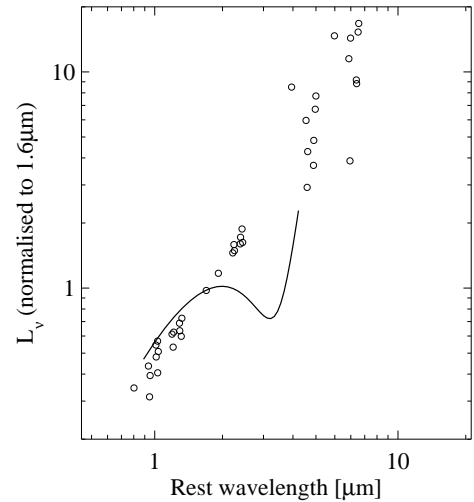}
\caption{The rest-frame near-infrared SED of our sample of $S_{8}/S_{4.5}>2$ SMGs (open cicles) compared with a fourth-order polynomial fitted to $S_{8}/S_{4.5}<2$ (SB-dominated) SMGs from Messenger et al.\ (in preparation). 
This simple comparison shows that on average our sample of $S_{8}/S_{4.5}>2$ SMGs are dominated by power-law continuum emission which swamps the stellar emission bump normally seen in typical SB-dominated SMGs.\label{fig:bump}}
\end{figure}

Although deeply-embedded H{\sc ii} regions could potentially be contributing to the hot dust continuum that we here associate with an AGN contribution (making the fractions quoted above merely upper limits to the AGN contribution), we argue below that the dust heated by stars is likely a minor contribution.  In Fig.~\ref{fig:bump} we have plotted our 3.6--24\,$\mu$m photometry in the rest-frame to search for any evidence of the H-opacity minimum (1.6\,$\mu$m stellar bump), as traced by the 4.5 and 5.8\,$\mu$m channels given our source redshifts, which is normally visible in typical SF SMGs (e.g.~\citealt{Hainline09}; Messenger et al.\ in preparation).  The rest-frame broad-band photometry of our sample of SMGs reveals a strong power-law shape from 1--3\,$\mu$m.  This exercise reveals that on average our sample of $S_{8}/S_{4.5}>2$ SMGs shows no 1.6\,$\mu$m stellar bump, whereas more typical SB-dominated SMGs ($S_{8}/S_{4.5}<2$) do (see also \citealt{Yun08}).  We thus interpret the $S_{8}/S_{4.5}>2$ observed color excess as due to hot dust emission (500--1000\,K) in the immediate vicinity of an active nucleus \citep{Laurent00}.  By extension, this AGN emission would also dominate at the longer wavelengths probed by the IRS spectra ($\lsim10$\,$\mu$m), causing the power-law shape that we observe.  Tentative evidence for AGN emission in these SMGs is apparent at radio wavelengths in the form of enhanced 1.4\,GHz emission compared to 610\,MHz data (see \S~\ref{sec:indices}).  We can directly test how likely it is that the AGN is producing the steep continuum emission in the mid-infrared spectra by examining the combination of the mid-infrared and X-ray data, as we now describe.

\begin{figure}
\epsscale{0.8}
\plotone{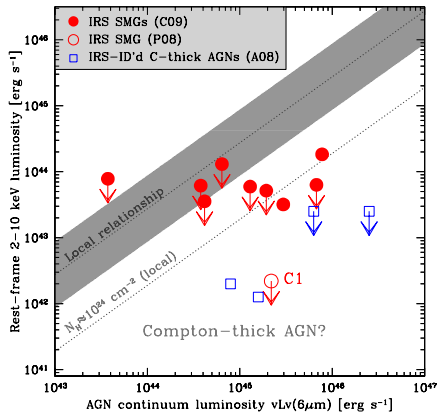}
\caption{Rest-frame 2--10\,keV luminosity derived from the X-ray versus 6\,$\mu$m luminosity derived from the mid-infrared power-law component fit (filled circles, where arrows indicate X-ray 3\,$\sigma$ upper limits). For comparison, we overplot \textit{Spitzer}-IRS $z\sim2$ AGN with optical spectroscopy from \citet{Alexander08} (open squares) and the SMG C1 from \citetalias{Pope08} (open circle) which have properties suggesting they are Compton-thick .  We have also overplotted the intrinsic (absorption corrected) X-ray--to--mid-infrared luminosity relationship for local AGN (\citealt{Lutz04}; the shaded region) and, assuming that it holds at high-redshift, the combination of the observed (not corrected for absorption) X-ray luminosities and 6\,$\mu$m luminosities of some of our objects suggests that they could be Compton-thick (see Fig.~4 of \citealt{Alexander08}; the lower dotted line refers to the observed luminosity ratio expected for a typical AGN that is absorbed by $N_\mathrm{H}\simeq10^{24}$\,cm$^{-2}$ using the model presented in \citealt{Alexander05}).  Note that C1 from \citetalias{Pope08} is also continuum-dominated in the mid-infrared, but it possesses a much lower $L_\mathrm{X}$ upper limit since it lies in a much deeper X-ray field.  
\label{fig:xvsir}}
\end{figure}

Using the mid-infrared continuum fits from \S~\ref{sec:analysis2}, we calculate rest-frame 6\,$\mu$m luminosities due to the AGN component and compare these with the X-ray luminosities (Table~\ref{tab:agn}).  Comparing the X-ray and mid-infrared data can potentially tell us about the intrinsic luminosity of the AGN and can reveal whether any of the sources are Compton-thick by comparing the estimated column densities to other samples of SMGs and local AGN. Following \citet{Alexander08}, in Fig.~\ref{fig:xvsir} we have plotted the rest-frame 2--10\,keV versus 6\,$\mu$m luminosities for our sample compared to AGN in the local Universe and to other SMGs.  Assuming the X-ray--to--mid-infrared luminosity relationship for local AGN \citep{Lutz04} holds at high-redshift, the combination of the intrinsic X-ray luminosities and 6\,$\mu$m luminosities of some of our objects suggests that they may be Compton-thick (i.e.~4 have upper limits close to the lower dotted line in Fig.~\ref{fig:xvsir}).  While the X-ray data are not sufficiently deep (cf.~2\,Ms $Chandra$ data in GOODS-N, e.g.~\citealt{Alexander08}) to conclusively determine whether or not the AGN are Compton-thick, the mid-infrared spectra show that AGN are dominating in these sources (in the form of strong continuum emission), and the non-detections are consistent with sources with being absorbed at X-ray energies (i.e.~$N_\mathrm{H}\gsim10^{22}$\,cm$^{-2}$).  Indeed since 7/8 SMGs are also optically faint with $R>23$, the majority of the sample appear to be obscured to some degree.  Compared to other SMGs, most of our objects are consistent with typical moderate-luminosity AGNs in ULIRGs/SMGs, while two of our objects (AzLOCK.01 and AzLOCK.10) are more consistent with extreme AGN-dominated types.  

\subsubsection{Mid-infrared continuum (AGN) properties}

In light of the results in \S~\ref{sec:rel}, we assume from now on that the continuum components of the fits are due to AGN activity, and we compare the mid-infrared AGN properties of our sample with more typical SB-dominated SMGs.  In general, the continuum portion of our fits from \S~\ref{sec:analysis} reveals that our SMGs display more strongly rising mid-infrared continuum when compared to more typical PAH-dominated SMGs (which comprise $\sim80$\% of the SMG population; \citetalias{Pope08}; \citetalias{KMD09}).  In addition, while we do not see strong evidence for silicate absorption at $\sim9.7$\,$\mu$m in our IRS spectra (which is often seen in AGN-dominated source), it cannot be ruled out since the S/N of the continuum between 9--10\,$\mu$m is not high enough.  Notably, \textit{all} of the best-fitting models to our sample require some level of `additional' continuum component compared to typical SMGs.  Unfortunately, the data are of relatively low S/N and possess insufficient wavelength coverage to provide robust measurements of individual object spectral slopes.  Nevertheless, for completeness, we find a range of $\alpha=0$--4.6 (uncorrected for extinction, and assuming the convention $S_{\nu}\,\propto \nu^{-\alpha}$) with a mean and median of 1.6 and 1.8, respectively.  In contrast, \citetalias{KMD09} report a median $\alpha$ of 1.05 for their radio-detected sample of (mostly) PAH-dominated SMGs. In addition, \citetalias{Pope08} found that 8/13 of their sample of more typical SB-dominated SMGs require a negligible level of continuum emission (and more often none) compared to the PAH template that was fitted.

\subsubsection{PAH properties}

Here we present the properties of the PAH features measured in \S~\ref{sec:analysis1} (see Table~\ref{tab:lines}), a key signature of luminous SF galaxies. We have tried to achieve the best PAH line measurements possible for this sample in order to compare our results with those of more typical SB-dominated samples of SMGs, and to other galaxy populations such as ULIRGs and AGNs, in order to yield insight into the physics of these systems.  However, many of our PAHs are swamped by continuum emission and so this makes a fair comparison with other samples difficult, given the intrinsic differences in the spectra, the systematic differences in the way lines are measured for the different sub-samples (e.g.~how extinction is taken into account), and the large measurement uncertainties in the derived quantities (see e.g.~\citetalias{Pope08} for a discussion).  Nevertheless, qualitatively we find that within the systematic uncertainties our 6.2\,$\mu$m PAH luminosities\footnote{We limit our comparison of PAH line fluxes to the 6.2\,$\mu$m feature since the 7.7\,$\mu$m feature is blended with other PAH features (8.6\,$\mu$m, for example) and the 9.7\,$\mu$m extinction trough which further complicates comparisons between different samples.} (median $2.7\times10^{10}\,L_\odot$) are towards the top end of SB-dominated SMGs studied by \citetalias{Pope08} and \citetalias{KMD09}, who find a spread covering about $\sim0.5$--3.5$\times10^{10}\,\mathrm{L}_\odot$ in their samples.  The errors on the EWs are currently too uncertain to determine if there is a genuine spread in the EW of SB/SMGs compared with our sample, although we see the trend noted by \citetalias{Pope08} and \citetalias{KMD09}, that the SMGs in our sample with the largest AGN fractions have the smallest 6.2 and 7.7\,$\mu$m EWs.  Next, we discuss the similarities and differences in the bulk PAH properties between this sample and others by constructing a composite spectrum of our AGN-dominated SMGs (see \S~\ref{sec:decomp}).

\subsection{Composite for AGN-dominated SMGs}\label{sec:decomp}

The spectral decomposition has revealed that some of the SMGs in our sample look more similar to those in \citetalias{Pope08}, with low AGN contributions to the mid-infrared, while some of our SMGs clearly look more like the rarer AGN-dominated Compton-thick objects like C1 from \citetalias{Pope08}. 
In order to compare the ensemble of properties of AGN-dominated versus SF-dominated SMGs and other galaxy populations, we have combined the IRS spectra of AGN-dominated ($>50$\% continuum contribution in the mid-infrared) SMGs from all available samples and created a higher S/N composite spectrum.  Using this composite we can search for and characterize faint features.  

\begin{figure}
\epsscale{0.8}
\plotone{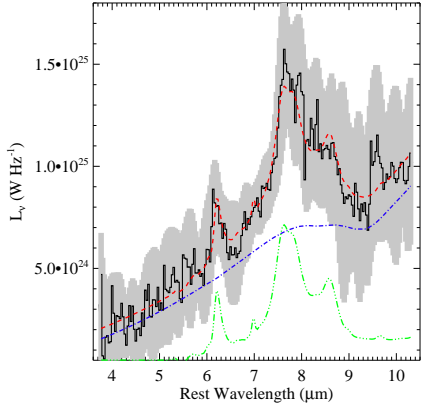}
\caption{Composite IRS spectrum of 10 mid-infrared AGN-dominated SMGs
(see text). The shaded region shows the 1$\sigma$ uncertainty. The red dashed line is the best-fitting model (see \S~\ref{sec:analysis2}), which is composed of a PAH template (green dot-dot-dot-dashed line) and a power-law component with extinction applied (blue dot-dashed line), revealing a 72\% AGN fraction and an unabsorbed power-law component slope of $\simeq2$ (see \S~\ref{sec:decomp}).}\label{fig:comp}
\end{figure}

%
%
%
%
%
%

\begin{figure}
\epsscale{0.8}
\plotone{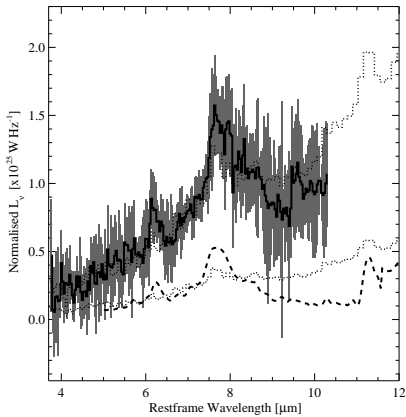}
\caption{Composite IRS spectrum of 10 mid-infrared AGN-dominated SMGs
(black histogram, with the shaded region representing the $1\,\sigma$ uncertainty).  We have overlaid the SB-dominated SMG composite from \citetalias{Pope08} (dashed curve), and note that the AGN-dominated SMG composite is $\simeq5$ times more luminous at $\sim7$\,$\mu$m than the SB-dominated SMG composite).  We have also overplotted the $\tau_{9.7}<1$ weak PAH composite of 17 24\,$\mu$m-selected ULIRGs (dotted histograms) from \citet{Sajina07} twice, once renormalised at $\sim7$\,$\mu$m to match the AGN-dominated composite and also separately to match the SB-dominated composite, in order to highlight the relative similarities and differences between the 3 composites.  This comparison reveals that AGN-dominated SMGs appear to have features broadly in between that of SMGs and ULIRGs.}\label{fig:compare}
\end{figure}
%
%
%
%
%
%

The composite is comprised of ten SMGs including our five mid-infrared AGN-dominated SMGs,  C1 (82\%) and GN04 (61\%) from \citetalias{Pope08}, as well as SMMJ105238.30, SMMJ123553.26, and SMMJ163650.43 from \citetalias{KMD09}\footnote{Since \citetalias{KMD09} do not explicitly provide individual mid-infrared AGN fractions, we have re-derived them here, following the procedure in \S~\ref{sec:analysis2}, and find AGN fractions of 94, 85, and 69\%, respectively, fulfilling our `AGN-dominated' criterion.}.  
Since the individual spectra span a wide dynamic range in rest-frame luminosity, we normalized them all to the median $\sim7\,\mu$m luminosity of the sample\footnote{We normalise the spectra to the median total luminosity between rest-frame 6.5--7.5\,$\mu$m, where there is a lack of PAH features or absorption, in order to minimize the potential bias.}.
We calculate the standard deviation of the sample as a measure of the error on the composite spectrum, which dominates over the uncertainty from individual flux elements.   We restrict the final SMG composite spectrum wavelength coverage to areas where there are $>3$ data files, resulting in a coverage of $\sim$4--10$\,\mu$m in the rest-frame (see Fig.~\ref{fig:comp}).

We perform a spectral decomposition on the AGN composite, following the method outlined in \S~\ref{sec:analysis2}, revealing a 72\% AGN fraction and an unabsorbed power-law component slope of $\simeq2$.  The amount of extinction required is $\tau_{9.7}\simeq0.5$, but is not well constrained (as mentioned previously) since the spectrum does not extend much past 9.7\,$\mu$m (but by eye, it seems that the slope is flatter due to extinction at longer wavelengths).

In Fig.~\ref{fig:compare} we compare our mid-infrared AGN-dominated SMG composite to the more typical PAH-dominated SMG composite of \citetalias{Pope08} and to a composite of $z\sim2$ bright ($S_{24}>0.9$\,mJy) red 24\,$\mu$m-selected ULIRGs from \citet{Sajina07}, which are mainly weak-PAH emitters and AGN-dominated sources.  First of all, Fig.~\ref{fig:compare} demonstrates that our AGN-dominated composite is $\simeq5$ times brighter at $\sim7$\,$\mu$m than the SB-dominated SMG composite, and we have not renormalized them to the same luminosity in order to highlight this difference.  Given that the observed mid-infrared spectra and fluxes are so similar between these two samples, about a factor of $\simeq1.5$--2 can be accounted for simply by the apparently higher average redshift of the AGN-dominated combined sample (median $z=2.56$) compared with the SF-dominated SMG sample (median $z=1.98$), and we discuss the possibility of selection effects further in \S~\ref{sec:discuss}.  Fig.~\ref{fig:compare} reveals that on average these SMGs have proportionately stronger/steeper mid-infrared continuum emission than more typical SF SMGs, as well as PAH emission at 6.2, 7.7, and 8.6\,$\mu$m, as well as a hint of silicate absorption at $\sim9.7$\,$\mu$m (although the data are very noisy in this region of the spectrum).  At $>9$--10\,$\mu$m  the average AGN-dominated SMG is more similar to the low opacity ($\tau_{9.7}<1$) $24\,\mu$m-selected ULIRGs from \citet{Sajina07} than to the SMG composite.  In summary, in terms of their mid-infrared spectra, AGN-dominated SMGs broadly appear to have features somewhere in between SMGs (strong SF activity indicated by PAH emission) \textit{and} 24\,$\mu$m-selected ULIRGs (ongoing SF with a relatively strong AGN component).

\subsection{Full SED Fits}\label{sec:fits}

\begin{figure*}
\epsscale{0.8}
\plotone{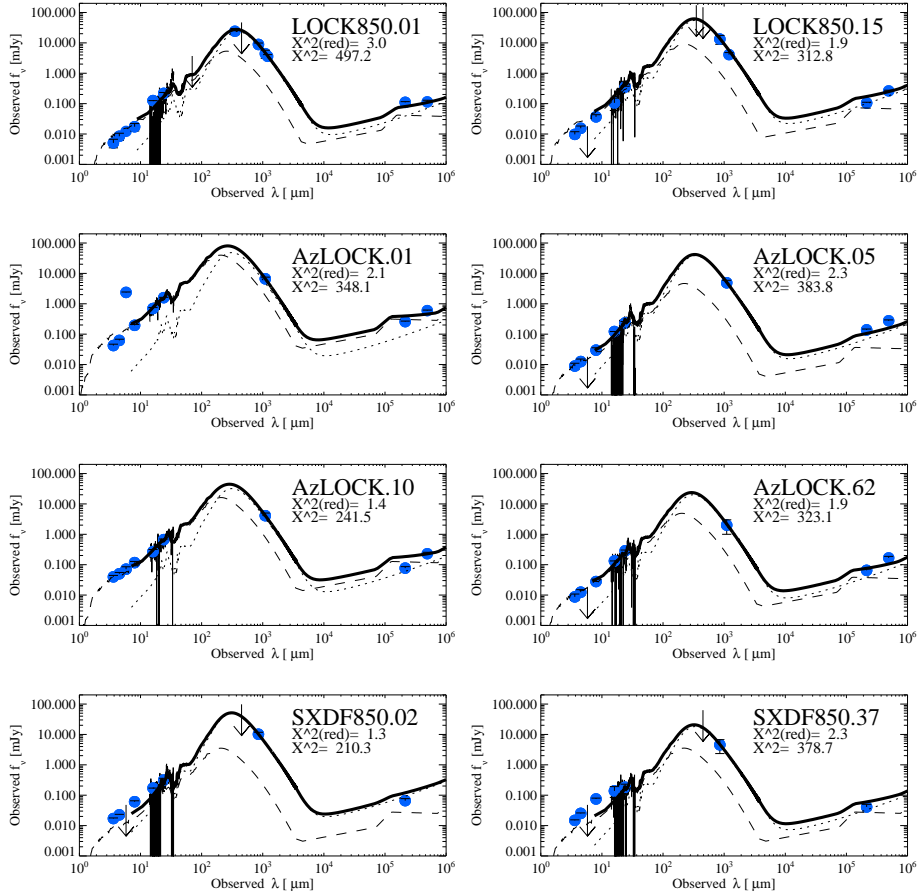}
\vspace{0.5in}
\caption{Mid-infrared--to--submm SEDs of our sample of SMGs.  The dark solid curve shows the best fitting linear combination of an SB component (the SMG starburst composite SED from \citetalias{Pope08} comprised of their SMG SB composite spectrum spliced with the best-fitting modified CE01 template, dotted curve) and an AGN component (a Mrk\,231 template, dashed curve) to the IRS spectrum (histogram) and 850 or 1100\,$\mu$m photometry.  Source names are given in each panel (see Table~\ref{tab:obs}), as well as the best-fit $\chi^{2}$ values.  We also include 350\,$\mu$m photometry \citep{Coppin08a} in the fitting for LOCK850.01.  We note a significant degradation in the quality of the fits in the mid-infrared if the additional AGN component is not included, indicating that this subset of SMGs is generally well described by the SEDs of typical SMGs, but with an additional AGN component required.  For completeness, we overlay 3.6--24\,$\mu$m and 610\,MHz and 1.4\,GHz photometry (see Table~\ref{tab:targets}) as well as 70 and 450\,$\mu$m 3\,$\sigma$ upper limits from \citet{Hainline09} and \citet{Coppin06}, respectively, to demonstrate that these data are roughly consistent with the template fitting, even though they are not used explicitly.\label{fig:sed}}
\end{figure*}

We have established mid-infrared AGN-dominance for the majority of our SMG sample.  However, it is not clear if the AGN activity will continue to produce a dominant fraction of the total infrared luminosity, since it is expected that the AGN will become less important bolometrically than SBs when extrapolating from the mid-infrared to longer wavelengths (\citealt{Tran01}; \citetalias{Pope08}).  Here we investigate the effect of the AGN emission to the total infrared luminosity when longer wavelength (submm) emission near the peak of the bolometric energy output of the galaxy is taken into account.  

To begin with, we fit the mid-infrared spectrum and submm/mm data (but not the radio data; see \S~\ref{sec:indices}) for each source with the full SF-SMG composite SED from \citetalias{Pope08}, which describes typical SMGs well (and includes a $\simeq30$\% AGN contribution).  Visually, all of the fits look poor and unsurprisingly have values of $\chi^{2}_\mathrm{red}\gg1$, meaning that the assumed model does not provide a very good description of the data.  Since we already have evidence from the mid-infrared fitting in \S~\ref{sec:rel} supporting a significant continuum/AGN contribution over typical SMGs in many cases, we repeat the fitting with a linear combination of the SF-SMG composite \textit{and} an additional AGN component represented by Mrk\,231, as described in \S~\ref{sec:analysis3}.  Visually the fits improve drastically, and we find an improvement in $\chi^{2}_\mathrm{red}$ values in \textit{every} case, typically of $\Delta\chi^{2}_\mathrm{red}\simeq$3 (though as large as $\sim10$ for AzLOCK.10 and $\sim60$ for AzLOCK.01), supporting the inclusion of the additional AGN template into the total SED model (see Fig.~\ref{fig:sed}).  

A wider range of templates and/or models could be explored in principle, however many complications could arise and make the results difficult to interpret. Since we wish to compare our decomposition directly with that of more typical SB-dominated SMGs (\citetalias{Pope08}), the most natural choice for the AGN component is the SED of Mrk\,231 since this is also what \citetalias{Pope08} used.  Mrk\,231 is a well-studied local ULIRG with strong evidence for an emerging AGN.  For example, \citet{Genzel98} conclude from high-excitation line measurements that Mrk\,231 is likely to be powered mainly by an AGN, and more recently, \citet{Armus07} find a mid-infrared AGN fraction of $\sim100$\%.  Although Mrk\,231 appears to be AGN-dominated in several wavebands it could also harbor some amount of dusty SF which would bias our bolometric AGN fractions upward.  In order to verify this possibility we repeat the SED fitting with the nuclear spectrum of the well-known local AGN NGC\,1068 (R. Chary priv. comm.; \citealt{Genzel98}; \citealt{Lutz00}) in place of Mrk\,231 and, although the relative shapes of their dust spectra indicate that NGC\,1068 is cooler than Mrk\,231, the fitting yields the same AGN fractions to within a few percent.  We also explore SF/AGN combinations using M82 as the SB component (instead of the \citetalias{Pope08} composite) and find that while it reproduces the mid-infrared PAH features well, M82 catastrophically fails to fit the submm/mm photometry, as expected since M82 has a much warmer dust temperature than SMGs.   
A more extensive comparison using a suite of many-parameter models (e.g.~\citealt{SK07}) is beyond the scope of our simple approach given the sparsity of the data, although could be attempted in the near future as the data quality and SED coverage could be improved with upcoming \textit{Herschel} snd \textit{JWST} surveys, for example.

Using these fits, we have calculated values of total $L_\mathrm{IR}$ as well as $L_\mathrm{SB}$ and $L_\mathrm{AGN}$ for the SF and AGN components separately and give these in Table~\ref{tab:lines} (see \S~\ref{sec:analysis3}).  The fractional contribution of AGN to the total $L_\mathrm{IR}$ in our sample ranges from 17--63\%, with evidence for only two SMGs being bolometrically dominated by an AGN ($>50$\% maximal AGN contribution -- AzLOCK.01 and AzLOCK.10 -- which also have the strongest AGN components in the mid-infrared).  Combining our five SMGs with a $>50$\% AGN-contribution with C1 and GN04 from \citetalias{Pope08}, the average AGN contribution to the bolometric luminosity, $L_\mathrm{IR}$, is  $\simeq40$\%, compared with 15\% for typical star-formation dominated SMGs from \citetalias{Pope08}.  We thus find a median $L_\mathrm{SB}$ for our sample of $6.5\times10^{12}$\,L$_\odot$, which translates into a median SFR for the sample of $\simeq1000$\,M$_\odot$\,yr$^{-1}$, similar to typical bright SB-dominated SMGs with $5\times10^{12}$\,L$_\odot$ (\citetalias{Pope08}).  Our sample thus has a similar $L_\mathrm{IR}$ to typical SB-dominated SMGs (only $\sim1.5$ times more compared to \citetalias{Pope08}), and thus it appears that our SMGs are from a similar SB luminosity class with similar SFRs, but that our sample has proportionately higher AGN content in the mid-infrared.  We discuss the implications of these results and how they might fit into an evolutionary context with other SMGs in \S~\ref{sec:discuss}.

We have only fit the SEDs to the mid-infrared spectra and the submm data, but note that extrapolation of the best-fitting SEDs does generally fit the observed IRAC and 610\,MHz photometry of our SMGs well, with some amount of scatter at 1.4\,GHz (see Fig.~\ref{fig:sed}).  Radio frequencies of 1.4\,GHz are sensitive to synchrotron radiation from relativistic electrons from supernovae (and hence recent star-formation; \citealt{Condon92}), although an AGN can also produce comparable synchrotron emission by jets and radio lobes, leading to ambiguities in the cause of the mechanism at $\mu$Jy levels in the absence of other information. Since we know that these SMGs are harboring AGN, it is quite possible (and likely) that the cause for the disagreement between the SED fits and the 1.4\,GHz data is due to an AGN which we now investigate by examining their measured 1.4\,GHz--610\,MHz spectral indices for clues.

\subsubsection{Radio Spectral Indices}\label{sec:indices}

\citet{Ibar09b} calculate 1.4\,GHz--610\,MHz radio spectral indices based on radio maps from \citet{Ibar09} using a fixed-to-beamsize source extraction, and they find a mean 1.4\,GHz--610\,MHz radio spectral index of $\alpha_{610}^{1.4}\,= -0.75\pm0.06$ ($S_\nu \propto \nu^{\alpha}$) for the bulk of SMGs in the LH, consistent with optically thin synchrotron emission.  The spectral indices for those of our SMGs which lie in the LH have a range of $-1.32<\alpha<-0.05$ (see Table~\ref{tab:targets}).  Five of these LH SMGs have steep radio spectra, $\alpha_{610}^{1.4}\,\lsim -0.84$, while one has a self-absorbed flat spectrum, all of which are clearly deviant from typical SMGs.  This difference seen in this subset of SMGs is suggestive that the radio emission has a different origin in these systems.  Since we know from the mid-infrared spectral signatures that these SMGs likely harbor AGN, it seems reasonable to assume that the AGN is affecting the radio emission, although other environmental causes could be playing a role \citep{Ibar09b}.  In order to quantify this we calculate the logarithmic ratio of the rest-frame infrared and radio luminosities for SMGs using the $q_\mathrm{L}$ parameter from \citet{Kovacs06}, and find a range of $q_\mathrm{L}$ values from 2.0--2.5, with a median value of $2.21\pm0.15$ compared to $2.14\pm0.07$ for radio-detected bright SMGs.  Although there is no evidence for an offset in $q_\mathrm{L}$ compared to \citet{Kovacs06}, the spectral indices suggest that the radio emission could be contaminated by radio-quiet AGN emission, providing some justification for omitting these data from the fitting.

\section{Discussion}\label{sec:discuss}

These new IRS spectra of AGN-dominated SMGs have helped to confirm the predictive power of the $S_{24}$/$S_{8.0}$--$S_{8.0}$/$S_{4.5}$ diagram.  Our sample was selected to lie off the expected SB sequence in the diagnostic color-color plot (Fig.~\ref{fig:selection}), and we have confirmed that the continuum emission dominates over PAH emission in the mid-infrared for 5/8 of our sources, which we interpret as due to a powerful AGN.  The majority of our SMGs possess stronger/steeper mid-infrared continuum emission than more typical SB-dominated SMGs, indicating that there is a continuous trend in AGN content in SMGs -- from low-luminosity AGN to extremely luminous AGN.  As with all color-selection methods, the IRAC color selection is merely a guide to singling out the most powerful AGN, and some level of contamination from SB systems is to be expected.  In particular, SF systems at $z\lesssim1$ could have $S_{8}/S_{4.5}>2$, since the 4.5 and 8\,$\mu$m channels are not yet climbing up the restframe 1.6\,$\mu$m stellar bump.  For example, the most discrepant data point on the AGN track is an SMG at $z=0.689$ from \citet{Hainline09} which is clearly SB-dominated in the IRS spectrum (\citetalias{KMD09}).  The color cut is thus appropriate for separating mid-infrared SB- and AGN-dominated SMGs from $z\simeq1$--4.  However, since most SMGs lie within $1<z<4$ \citep{Chapman05}, negligible contamination is expected on either side of the color cut boundary.  In light of the IRS results of \citetalias{KMD09}, who observed a large fraction of the submm-bright radio-selected sample of \citet{Chapman05} which have optical spectroscopic redshifts, we advocate adjusting the selection slightly to $S_{8}/S_{4.5}\gtrsim1.65$ to account for the AGN-dominated SMGs seen in the \citetalias{KMD09} sample.  Even then, we still see some small amount of scatter to either side of this line (see Fig.~\ref{fig:selection}).  Combining all the IRS-observed blank-field SMG samples, we find that $\simeq15$\% of SMGs are mid-infrared AGN-dominated systems.  Our full SED fitting reveals that 2/8 (25\%) of SMGs in our sample with AGN-dominant mid-infrared emission are likely bolometrically dominated by an AGN (in the far-infrared).  This implies that overall $\sim5$\% of blank-field SMGs are likely to have a dominant contribution to their $L_\mathrm{IR}$ due to AGN activity.

Now that we have confirmed that the majority of our sample is comprised of star-forming SMGs with an excess of AGN emission compared to typical SMGs, we can interpret this result within the framework of the proposed evolutionary sequence which links our AGN-dominated sample of SMGs to typical SMGs.  However, before placing these results in any wider context, we must first take stock of any potential selection effects and how they could bias our interpretation. 

\subsection{Selection Effects}\label{sec:bias}

The eight targets in our color-selected sample have 850$\,\mu$m and 24$\,\mu$m flux densities of 4--13$\,$mJy\footnote{Note that we have converted the AzTEC 1.1\,mm fluxes to 850\,$\mu$m fluxes by multiplying by a factor of 1.8, according to the average $S_{850}/S_{1100}$ color ratio observed for SMGs (e.g.~\citealt{Chapin09}).} and 180--1600$\,\mu$Jy, with median values of 9\,mJy and 310\,$\mu$Jy, respectively (see Table~\ref{tab:targets}).   Note that our targets have a similar 24\,$\mu$m flux density distribution as the whole SHADES SMG sample (i.e.~our color selection has not biased the sample to preferentially brighter or fainter 24\,$\mu$m SMG counterparts).   Thus, our IRAC color-selected subset of SMGs studied here should be representative of $\sim\!15$\% of the SMG population discovered in existing blank-field 850$\,\mu$m imaging. 
The primary submm/mm-selection uniformly identifies high-redshift galaxies with $L_\mathrm{SB}>10^{12}\,\mathrm{L}_\odot$ and corresponding SFRs of $\sim1000\,\mathrm{M}_{\odot}$\,yr$^{-1}$.  The AGN-dominated SMG sample we have focussed on here is similar to typical SMGs in terms of $L_\mathrm{SB}$, which allows for a fair and direct comparison between the two sub-samples of SMGs.  However, there is some evidence that our sample lies at a slightly higher average redshift than typical SMGs, which merits some discussion.

The redshift distribution of our color-selected sample of eight objects ranges from 2.5--3.4, with a median of 2.8.  This is noticeably shifted to higher redshifts compared to the \citetalias{Pope08} and \citetalias{KMD09} samples, which have a median redshift of 2.0, more consistent with the known redshift distribution for radio-identified SMGs with $S_{850}>5$\,mJy ($z\sim2.2$; \citealt{Chapman05}).  The bias in our redshift distribution to earlier epochs is likely a submm selection effect, which tends to pick out warmer galaxies on average at higher redshift (e.g.~\citealt{Pope06}; \citealt{Coppin08a}; \citealt{Dye09}).  The subsequent IRAC color-selection imposed could be effectively singling out those higher-redshift SMGs which possess warm dusty tori heated by a proportionately larger AGN.  This trend was also noted by \citetalias{KMD09} and \citet{Watabe09}, who found that `typical' SMGs show an increase in mid-infrared continuum strength (and less prominent PAHs) with increasing redshift.  One way to verify this trend is to measure the flux near the far-infrared dust peak of the SED in order to constrain the dust temperatures in our $S_{8}/S_{4.5}>2$ sample and compare them to the dust temperatures of more typical SMGs with $S_{8}/S_{4.5}<2$, over a wide redshift range.  Facilities such as \textit{Herschel} and SCUBA-2 will be able to help address this issue for this sample, although some progress could be made now using BLAST which has already measured the rest-frame IR peaks of hundreds of SMGs in other fields where \textit{Spitzer} IRAC data exist (\citealt{Devlin09}; \citealt{Dye09}).  However, it is unlikely that the average redshift difference (equivalent to $\simeq1$\,Gyr in cosmic time) will have a large effect on the main conclusions, since even SMGs at $z\sim4$--5 seem to show a mix of starburst and AGN signatures/SB properties similar to SMGs at $z\sim2$ (e.g.~\citealt{Coppin09}).

\subsection{Where do our AGN-dominated SMGs fit into the evolutionary sequence?}\label{sec:evol}

For the purposes of this discussion, we adopt the proposed evolutionary sequence first proposed by \citet{Sanders88} and test the scenario where the SMG population evolves through a primarily starburst phase, onto a subsequent submm-detected AGN-phase, into a submm-undetected QSO and finally a passive elliptical.  \textit{Spitzer}-IRS data has tentatively identified some local ULIRGs being in key stages in this lifecycle \citep{Farrah09b}.  Now we ask -- \textit{are the observed properties of typical SMGs and the AGN-dominated SMG sample studied here consistent with this picture in the high-redshift universe, in terms of their space densities and other observed properties?}

$S_{850}>4.5$\,mJy SMGs have surface densities of $\simeq600\,\mathrm{deg}^{-2}$ \citep{Coppin06} or space densities of $\simeq10^{-5}$\,Mpc$^{-3}$ for $z=1$--3 (\citealt{Chapman05}; \citealt{Wall08}).  The combination of this IRS work and other samples of SMGs has uncovered a mid-infrared presence of appreciable heating of a dust torus by an AGN for approximately 15\% of the SMG population at this epoch.  Assuming that all bright SMGs go through a subsequent AGN phase then the observed IRS-confirmed AGN-detection fraction merely represents a `duty cycle'.  The current total lifetime estimates from modelling SMGs are $\sim100$--300\,Myr (\citealt{Swinbank06}), and thus the inferred submm-bright AGN-dominated phase caught by the mid-infrared would be occuring during the last $\sim15$--45\,Myr of an SMG lifetime.  These relative lifetimes and duty cycle are consistent with the merger simulations of \citet{Springel05}, where the tidal forces from galaxy mergers trigger a nuclear SB, fuelling a rapid growth of BHs before the gas has been consumed.  The BH accretion rate grows throughout the peak phase of SF in the merger system (with the peak phase lasting $\approx100$\,Myr in these simulations), peaking $\approx50$\,Myr later than the SFR peak due to the delayed action of AGN feedback on the gas and manifesting itself as a luminous QSO for $\approx50$\,Myr total.  In the \citet{Springel05} simulations, the SMG phase is in principle both an SB and an AGN, but it is only during the final evolutionary stage that the remnant would be visible as an AGN, with outflows removing the surrounding gas and dust which heavily obscures the AGN at the beginning of the burst.  Tentative evidence for this has been found in typical SMGs since their estimated $M_\mathrm{BH}/M_\mathrm{gal}$ ratio is lower than found in $z\sim2$ QSOs (\citealt{Alexander08a}; \citealt{Coppin08b}; \citealt{Peng06}) 

Recently, \citet{Dey08} proposed that Dust Obscured Galaxies (DOGs; $S_{\nu}(24\,\mu\mathrm{m})/S_{\nu}(R)\gtrsim1000$ and $S_{24}>300$\,mJy, with surface densities similar to $>6$\,mJy SMGs) could represent a brief evolutionary phase between SMGs and less obscured QSOs or galaxies, which is supported by the finding that $\sim30$\% of SMGs (including the AGN-dominated SMGs) meet the DOG criteria \citep{Pope08b}.  Interestingly, we note that virtually all (5/7 with $R$-band coverage) of our targets here fulfill the DOG criteria, confirming that the majority of AGN-dominated SMGs also overlap with the DOG population. 

While these arguments are mainly circumstantial, we have shown that the AGN-dominated SMGs at least fit into the evolutionary picture of \citet{Sanders88} in terms of the inferred duty cycles.  While the statistical approach to duty cycles and relative lifetimes can be useful, in-depth studies of supposedly individual SMGs often reveal a complicated picture involving multiple components (e.g.~\citealt{Ivison08}).   Measuring gas masses with the IRAM Plateau de Bure Interferometer or the Atacama Large Millimeter Array, inferring gas depletion timescales, and estimating SMBH masses using near-infrared spectroscopy for the AGN-dominated sample studied here (see e.g.~\citealt{Alexander08a}; \citealt{Coppin08b}) would allow us to put further constraints on the SMG evolutionary sequence we are exploring.

\section{Conclusions}\label{sec:concl}

{\it Spitzer} IRS spectroscopy has been obtained for a sample of eight SMGs brighter than 200$\,\mu$Jy at 24$\,\mu$m and selected to possess observed IRAC colors of $S_{8}/S_{4.5}>2$, in order to investigate the level of AGN dominance in the SMG population.  Our conclusions are as follows:

1) Taking into account sources in the literature, we find a that a color-selection of $S_{8}/S_{4.5}>1.65$ is a better description overall for defining the boundary between SB and AGN-dominated SMGs, with a small amount of scatter across this division.  

2) We see signs of both SF and AGN activity in our sample of SMGs, with a continuous distribution of AGN fractions in the mid-infrared.  There are signs of SF in the form of PAH features in \textit{all} of the SMGs in our sample, from which we derive redshifts between 2.5--3.4, demonstrating the power of the mid-infrared to determine redshifts when the optical counterparts are too faint to study with current facilities.  

3) We find a median AGN contribution to the mid-infrared of 56\% in our sample of SMGs, indicating that overall, SMGs with $S_{24}>0.2$\,mJy and $S_{8}/S_{4.5}>2$ tend to have more dominant AGN-components in the mid-infrared than typical SMGs ($<30$\% mid-infrared AGN contribution). Extrapolation to the far-infrared reveals that the AGN is bolometrically important only in two of our SMGs (each with a $>75$\% mid-infrared AGN contribution).  This result suggests that significant contamination from AGN to the far-infrared luminosities affects $\lsim5$\% of the SMG population overall.  

4) To first order, these AGN range from being fairly low-luminosity types found in typical SMGs to more extreme cases that likely overlap with AGN-dominated $24\,\mu$m-selected ULIRGs in \citet{Sajina07}.  In any case, all our AGN appear to be obscured to some degree (at least five have low $L_\mathrm{X}/L_\mathrm{6\mu m}$ and six are optically faint with $R>23$) and a small subset of our sample may be obscured by Compton-thick material.

5) Our results are thus consistent with the \citet{Sanders88} evolutionary scenario, with all SMGs undergoing a `transitional' AGN-dominated phase with a duty cycle of $\simeq15$\%. Our sample of AGN-dominated SMGs could be at a slightly later stage of evolution than SF-dominated SMG systems, with the SF still occurring but where the AGN has now begun to heat the dust appreciably in the SMG as the BH undergoes a period of rapid growth.

\acknowledgments
This work is based on observations made with the \textit{Spitzer Space Telescope}, which is operated by the Jet Propulsion Laboratory (JPL), California Institute of Technology under a contract with NASA. The IRS was a collaborative venture between Cornell University and Ball Aerospace Corporation funded by NASA through the JPL and Ames Research Center.  Support for this work was provided by NASA through an award issued by JPL/Caltech.  This work is based in part on data obtained as part of the UKIRT Infrared Deep Sky Survey (UKIDSS).

The authors would like to thank an anonymous referee for comments which improved the paper.  KC acknowledges support from a UK Science and Technology Facilities Council fellowship.  AP acknowledges support provided by NASA through the {\it Spitzer Space Telescope} Fellowship Program, through a contract issued by the JPL, California Institute of Technology under a contract with NASA.  KMD is supported by an NSF Astronomy and Astrophysics Postdoctoral Fellowship under award AST-0802399.  DMA thanks the Royal Society and the Philip Leverhulme fellowship prize for generous support.  We also thank Ranga Chary for providing the full SED template of Mrk231 and the nuclear region of NGC1068, and to Laura Hainline and Jim Geach for useful discussions.


\begin{deluxetable}{llrrrrr}
\tabletypesize{\scriptsize}
\tablecaption{{\it Spitzer}-IRS observations of SHADES SMGs in the Lockman Hole (`LOCK') and the Subaru-$XMM$ Deep Field (`SXDF').}
\tablehead{
\multicolumn{1}{c}{IAU Name} & \multicolumn{1}{c}{Nickname} & \multicolumn{2}{c}{24\,$\mu$m Position (J2000)} & \multicolumn{3}{c}{Integration time}\\ 
& & \multicolumn{1}{c}{R.A.} & \multicolumn{1}{c}{Dec.} & \colhead{LL2\,($\times120$s)}   & \colhead{LL1\,($\times120$s)}   & \colhead{Total\,(hr)}
\label{tab:obs}
}
\startdata 
SHADES\,J105159+572423      & LOCK850.41-2  &  10\,51\,59.81 & 57\,24\,25.1  & 3$\times$6   & 2$\times$6  & 1.0 \\ 
AzTEC\,J105211.61+573510.7  & AzLOCK.62     &  10\,52\,11.85 & 57\,35\,10.5  & 14$\times$6  & 10$\times$6 & 4.8 \\
AzTE\,CJ105201.98+574049.3  & AzLOCK.01     &  10\,52\,01.92 & 57\,40\,51.5  & 2$\times$6   & 2$\times$6  & 0.8 \\ 
SHADES\,J021803--045527     & SXDF850.02    &  02\,18\,03.54 & --04\,55\,26.9 & 14$\times$6  & 11$\times$6 & 5.0 \\ 
AzTEC\,J105406.44+573309.6  & AzLOCK.10     &  10\,54\,06.83 & 57\,33\,09.1  & 3$\times$6   & 2$\times$6  & 1.0 \\ 
SHADES\,J105319+572110      & LOCK850.15    &  10\,53\,19.26 & 57\,21\,08.3  & 9$\times$6   & 7$\times$6  & 3.2 \\
SHADES\,J021724--045839     & SXDF850.37    &  02\,17\,24.41 & --04\,58\,42.0 & 31$\times$6  & 30$\times$6 & 12.2 \\
AzTEC\,J105403.76+572553.7  & AzLOCK.05     &  10\,54\,03.75 & 57\,25\,53.5  & 20$\times$6  & 16$\times$6 & 7.2 \\  
SHADES\,J105201+572443      & LOCK850.01    &  10\,52\,01.30 & 57\,24\,46.1  & 24$\times$6  & 20$\times$6 & 8.8 \\ 
\hline
Total time on-source & & & & & & 44.0 \\
\enddata
\end{deluxetable}

\clearpage

\begin{sidewaystable}
\tiny
\caption{Optical--to--radio photometry of the SMG 24\,$\mu$m counterparts}
\begin{tabular}{crrrrrrrrrrrrcrrr}

\multicolumn{1}{c}{Nickname}   & \multicolumn{1}{c}{$B$} & \multicolumn{1}{c}{$R$} & \multicolumn{1}{c}{$i'$} & \multicolumn{1}{c}{$z$} & \multicolumn{1}{c}{$J$} & \multicolumn{1}{c}{$K$} & \multicolumn{1}{c}{3.6\,$\mu$m} & \multicolumn{1}{c}{4.5\,$\mu$m} & \multicolumn{1}{c}{5.8\,$\mu$m} & \multicolumn{1}{c}{8\,$\mu$m} & \multicolumn{1}{c}{16\,$\mu$m} & \multicolumn{1}{c}{24\,$\mu$m} & \multicolumn{1}{c}{850 or 1100\,$\mu$m} & \multicolumn{1}{c}{1.4\,GHz} & \multicolumn{1}{c}{$\alpha_{1.4}^{350}$} & \multicolumn{1}{c}{$\alpha_{610}^{1.4}$} \\
\multicolumn{1}{c}{}   & \multicolumn{1}{c}{} & \multicolumn{1}{c}{} & \multicolumn{1}{c}{} & \multicolumn{1}{c}{} & \multicolumn{1}{c}{} & \multicolumn{1}{c}{} & \multicolumn{1}{c}{($\mu$Jy)} & \multicolumn{1}{c}{($\mu$Jy)}& \multicolumn{1}{c}{($\mu$Jy)} & \multicolumn{1}{c}{($\mu$Jy)} & \multicolumn{1}{c}{($\mu$Jy)} & \multicolumn{1}{c}{($\mu$Jy)} & \multicolumn{1}{c}{(mJy)} & \multicolumn{1}{c}{($\mu$Jy)} & \multicolumn{1}{c}{} & \multicolumn{1}{c}{}\\
\label{tab:targets}
AzLOCK.01       & -- & -- & -- & -- & $22.84\pm0.10$ & $21.12\pm0.04$ & $42.1\pm4.5$ & $62.1\pm5.8$ & $2362.2\pm79.9$ & $194.8\pm14.3$ & $697.5\pm51.5$ & $1587.3\pm43.6$ &  $6.6^{+0.9}_{-1.0}$  & $258\pm11$  & $0.69\pm0.03$ & -$1.01\pm0.07$ \\
SXDF850.02      & 26.12 & 25.33 & 25.26 & 24.96 & 24.40 & 22.54 & $17.0\pm0.8$ & $22.5\pm0.8$ & $<48$ & $60.9\pm4.5$ & $<171.0$ & $313.0\pm47.0$  &  $10.2^{+1.6}_{-1.6}$ & $66\pm11$ & $0.91\pm0.06$ & --  \\
AzLOCK.10       & $23.04\pm0.01$ & $22.65\pm0.01$ & $22.40\pm0.01$ & $22.23\pm0.01$ & $20.87\pm0.06$ & $20.16\pm0.02$ & $39.7\pm4.4$ & $50.1\pm5.2$ & $71.0\pm14.0$ & $116.6\pm11.1$ & $269.1\pm46.1$  & $668.4\pm29.5$  &  $4.1^{+0.9}_{-1.0}$  & $77\pm9$ & $0.83\pm0.06$ & -$1.32\pm0.20$ \\
LOCK850.15      & $26.54\pm0.12$ & $25.02\pm0.05$ & $25.24\pm0.09$ & $24.50\pm0.11$ & $>22.33$ & $22.25\pm0.12$ & $9.7\pm2.2$  & $15.4\pm2.9$ & $<18$  & $36.7\pm6.5$   & $105.4\pm46.0$ & $353.0\pm20.0$  &  $13.3^{+4.3}_{-5.0}$ & $105\pm6$ & $0.88\pm0.09$ & -$1.10\pm0.10$ \\
SXDF850.37      & $27.82\pm0.26$ & $26.87\pm0.16$ & $26.96\pm0.20$ & $26.49\pm0.32$ & $25.20\pm0.79$ & $23.25\pm0.21$ & $14.8\pm0.6$ & $25.2\pm0.9$ & $<48$ & $75.0\pm5.2$ & $138.0\pm53.8$  & $183.0\pm47.0$  &  $4.6^{+2.2}_{-2.6}$  & $41\pm9$ & $0.85\pm0.15$ & -- \\
AzLOCK.05       & $25.94\pm0.07$ & $25.46\pm0.08$ & $25.85\pm0.16$ & $25.18\pm0.21$ & $>22.33$ & $>22.93$ & $8.9\pm2.1$  & $12.5\pm2.6$ & $<18$ & $29.7\pm5.9$ & $<121.8$ & $235.0\pm24.1$  &  $4.9^{+1.0}_{-0.9}$  & $138\pm9$ & $0.75\pm0.05$ & -$0.84\pm0.11$ \\
LOCK850.01      & $>27.17$ & $25.83\pm0.11$ & $25.35\pm0.10$ & $24.89\pm0.16$ & $>22.33$ & $>22.93$ & $5.1\pm1.6$  & $8.4\pm2.2$  & $12.2\pm2.2$  & $17.3\pm4.7$ & $<125.7$ & $217.0\pm16.0$  &  $8.9^{+1.0}_{-1.0}$  & $110\pm6$ & $0.79\pm0.03$ & -$0.05\pm0.18$ \\
\\
\end{tabular}
\tablenotetext{\,}{Optical--to--radio multiwavelength magnitudes or flux densities of our sample of SMG 24\,$\mu$m counterpart targets, listed in order of redshift (see Table~\ref{tab:agn}).  Non-detections are indicated by the corresponding 5\,$\sigma$ point source sensitivity limit of the appropriate data set.  Columns 2--5 list the Subaru $BRi'z$ optical photometry (\citealt{Dye08}; \citealt{Furusawa08}; Mortier et al.\ in preparation), while columns 6--7 provide the $JK$-band photomertry from the DR3 UKIDSS release (\citealt{Lawrence07} and Warren et al.\ in preparation; see \S~\ref{sec:sample}). The optical/near-IR photometry is in AB magnitudes and has been measured in $2''$ apertures (with 5\,$\sigma$ upper limits given when the extracted photometry is fainter than these limits). The 3.6--8\,$\mu$m photometry for the LH SHADES, LH AzTEC, and SXDF SHADES sources are from \citet{Dye08} and \citet{Hainline09}, Egami et al. (priv. comm.), and SWIRE \citet{Lonsdale04} and SpUDS (Dunlop et al.\ priv comm.), respectively, and are listed in columns 8--11. New 16\,$\mu$m photometry data (this work) are listed in column 12.  The deboosted submm and mm photometry from \citet{Coppin06} and \citet{Austermann09}, respectively, are listed in column 14.  Columns 13 and 15 give the 24\,$\mu$m and radio flux densities from \citet{Ivison07}, \citet{Ibar09,Ibar09b}, and Arumugam et al.\ (in preparation).  The submm-to-radio spectral indices (350\,GHz-to-1.4\,GHz) are listed in column 16, with $S_{850}$ for the 1100\,$\mu$m AzTEC sources calculated assuming an $S_{850}/S_{1100}$ ratio of 1.8 (see \citealt{Chapin09}).  The 610\,MHz-to-1.4\,GHz spectral indices from \citet{Ibar09b} are given in column 17 and are available for the LH sources only. `--' indicates no coverage at the source position in this band.}
\end{sidewaystable}
\normalsize
\clearpage

\begin{deluxetable}{cccccc}
\tabletypesize{\scriptsize}
\tablecaption{Mid-infrared and X-ray derived properties.}
\tablehead{
\multicolumn{1}{l}{SMG ID} & \multicolumn{1}{c}{$z_\mathrm{IRS}$} & \multicolumn{1}{c}{Continuum} & \multicolumn{1}{l}{$\nu\,L_\mathrm{6\mu m}$} & \multicolumn{1}{l}{$S_X$(0.5--2\,keV)} & \multicolumn{1}{l}{$L_X$(2--10\,keV)} \\
\multicolumn{1}{l}{}       & \multicolumn{1}{c}{} & \multicolumn{1}{c}{[\%]} & \multicolumn{1}{l}{[$10^{45}\,\mathrm{erg\,s^{-1}}$]} & \multicolumn{1}{l}{[$10^{-16}\,\mathrm{erg\,cm^{-2}\,s^{-1}}$]} & \multicolumn{1}{l}{[$10^{43}\,\mathrm{erg\,s^{-1}}$]}
\label{tab:agn} }
\startdata
AzLOCK.62  & $2.48\pm0.03$  & 32 & 0.416      & $<5.23$               & $<3.55$ \\
AzLOCK.01  & $2.50\pm0.02$  & 71 & 6.76 (7.21)      & $<9.07$               & $<6.31$ \\
SXDF850.02 & $2.55\pm0.06$  & 10 & 0.0374     & $<10.95$              & $<7.81$ \\
AzLOCK.10  & $2.56\pm0.03$  & 77 & 2.96       & 4.4$^{a}$     & 3.19    \\
LOCK850.15 & $2.76\pm0.01$  & 62 & 1.94       & $<6.08$               & $<5.16$ \\
SXDF850.37 & $2.78\pm0.04$  & 51 & 0.641      & $<15.14$              & $<13.0$ \\
AzLOCK.05  & $2.82\pm0.02$  & 26 & 0.378 (1.84)     & $<6.94$               & $<6.15$ \\
LOCK850.01 & $3.38\pm0.02$  & 56 & 1.29       & $<4.78$               & $<5.94$ \\
\enddata
\tablenotetext{\,}{In column 2 we list the spectroscopic redshifts derived from the fitting procedure described in \S~\ref{sec:analysis1} for the PAH features in the mid-infrared spectra.  In column 3 we give the AGN contribution to the mid-infrared, expressed as a percentage of the total mid-infrared luminosity covered by the IRS spectra, as derived from the continuum power-law fits from \S~\ref{sec:analysis2}.  Column 4 gives the rest-frame 6\,$\mu$m luminosities based on the best-fitting power-law component, assuming that the AGN contributes to all of the emission at 6\,$\mu$m (with the extinction-corrected values given in parentheses where appropriate).  Column 5 gives the observed-frame X-ray fluxes of the detection and 3\,$\sigma$ upper limits for the non-detections, and column 6 gives the corresponding rest-frame X-ray luminosities, calculated using the X-ray flux and the IRS-derived redshifts; $\Gamma=1.4$ is used to make small K-corrections (see \S~\ref{sec:sample}).}
\tablenotetext{a}{The counterpart is 1$''$ away from the 24\,$\mu$m position.}
\end{deluxetable}

\begin{deluxetable}{crrrrrrrrr}
\tabletypesize{\scriptsize}
\tablecaption{PAH Luminosities, Equivalent Widths and Infrared Luminosities}
\tablehead{
\colhead{SMG ID} & \multicolumn{3}{c}{PAH luminosity ($10^{9}\rm{L_{\odot}}$)} & \multicolumn{3}{c}{PAH equivalent width$^{a}$ ($\mu$m)} & \multicolumn{3}{c}{$L_{\rm{IR}}^{b}$ ($10^{12}$L$_{\odot}$)} \\
                 &                              6.2$\,\mu$m & 7.7$\,\mu$m & 8.6$\,\mu$m & 6.2$\,\mu$m & 7.7$\,\mu$m & 8.6$\,\mu$m & SB & AGN & Total 
\label{tab:lines}
}
\startdata
AzLOCK.62  & $19.4\pm1.1$ & $82.7\pm2.3$ & $16.8\pm4.6$     & $0.62\pm2.30$ & $3.43\pm11.44$ & $0.77\pm2.48$ & 3.2 & 1.8 & 5.0 \\
AzLOCK.01  & $76.5\pm24.5$ & $387.1\pm11.1$ & $64.4\pm5.1$ & $0.26\pm0.22$ & $1.80\pm1.68$ & $0.34\pm0.33$ & 7.9 & 13.6 & 21.5 \\
SXDF850.02 & $15.8\pm9.4$ & $161.9\pm8.0$ & $22.9\pm11.5$   & $0.62\pm5.03$ & $6.61\pm37.52$ & $0.96\pm4.50$ & 8.3 & 1.6 & 10.0 \\
AzLOCK.10  & $30.1\pm14.9$ & $168.4\pm28.7$ & $26.4\pm11.2$ & $0.23\pm0.14$ & $1.59\pm0.84$ & $0.27\pm0.14$ & 5.5 & 5.9 & 11.3 \\
LOCK850.15 & $26.6\pm6.5$ & $125.2\pm4.2$ & $36.3\pm4.5$ & $0.26\pm0.17$ & $1.23\pm0.52$ & $0.36\pm0.13$ & 10.7 & 4.2 & 14.9 \\
SXDF850.37 & $12.6\pm8.1$ & $86.8\pm4.3$  & $8.9\pm8.5$     & $0.42\pm0.88$ & $4.63\pm12.19$ & $0.54\pm1.52$ & 3.6 & 1.6 & 5.2 \\
AzLOCK.05  & $36.5\pm8.7$ & $193.0\pm8.0$ & $2.8\pm4.5$     & $1.21\pm8.38$ & $6.58\pm31.27$ & $0.10\pm0.37$ & 7.8 & 2.3 & 10.1 \\
LOCK850.01 & $25.7\pm5.0$ & $96.9\pm31.9$ & ...          & $0.27\pm0.50$ & $0.59\pm0.44$ & ...           & 6.5 & 3.3 & 9.8 \\
\enddata
\tablenotetext{a}{The equivalent widths are all in the rest-frame. }
\tablenotetext{b}{Obtained from the best fitting linear combination of an SB component (the SMG starburst composite SED from \citetalias{Pope08} comprised of their SMG SB composite spectrum spliced with the best-fitting modified CE01 template) and an AGN component (a Mrk\,231 template) to the IRS spectrum and 850 or 1100\,$\mu$m photometry (see text).  The errors are typically $\lsim10$\% on each component.}
\end{deluxetable}

\appendix

\section{Notes on Individual Sources}\label{app:notes}

\subsection{LOCK850.1}\label{sec:lock1}

LOCK850.01 is the brightest SCUBA source detected in the LH region of the 8\,mJy Survey (\citealt{Scott02}; LE850.01) and SHADES \citep{Coppin06} and is also one of the brightest 1200\,$\mu$m sources detected in the \citet{Greve04} MAMBO survey (LH1200.005).  \citet{Lutz01} identify extended $K$-band emission with the PdBI 1.2\,mm-confirmed position, and using the combination of redshift estimates based on the photometric properties of the Extremely Red Object (ERO) counterpart they provide a `best guess' of $z=3$.  Further spectroscopic follow-up of LOCK850.01 was obtained through optical and near-infrared spectroscopy (\citealt{Simpson04}; \citealt{Chapman03}; \citealt{Chapman05}; \citealt{Blain04}; \citealt{Ivison05}), although no secure redshift was forthcoming from the optically faint ERO.  A galaxy $\sim3''$ away from the radio counterpart was also targeted and estimated to lie at $z=2.148$ based on absorption lines in the LRIS spectrum.  This redshift was tentatively assigned to LOCK850.01, since the likelihood of finding a $z\sim2$ galaxy so close to the SMG centroid by chance is slim and so was deemed likely to be associated with the optically-faint ERO (see \citealt{Ivison05} for a complete discussion).  Based on different photometric redshift techniques \citet{Aretxaga07} and \citet{Dye08} find $z_\mathrm{phot}=2.4^{1.1}_{0.2}$ (with a 90 per cent confidence range of 2.2--3.8) and $z_\mathrm{phot}=4.21$ (with a 90 per cent confidence range of 2.41--4.45), respectively.  Based on the presence of two PAHs in Fig.~\ref{fig:alex1}, $z=3.38\pm0.02$ is the $\chi^{2}$ best-fit solution and we adopt this redshift for LOCK850.01.  

\subsection{LOCK850.41-2}\label{sec:lock41}

\begin{figure}[h]
\epsscale{0.8}
\plotone{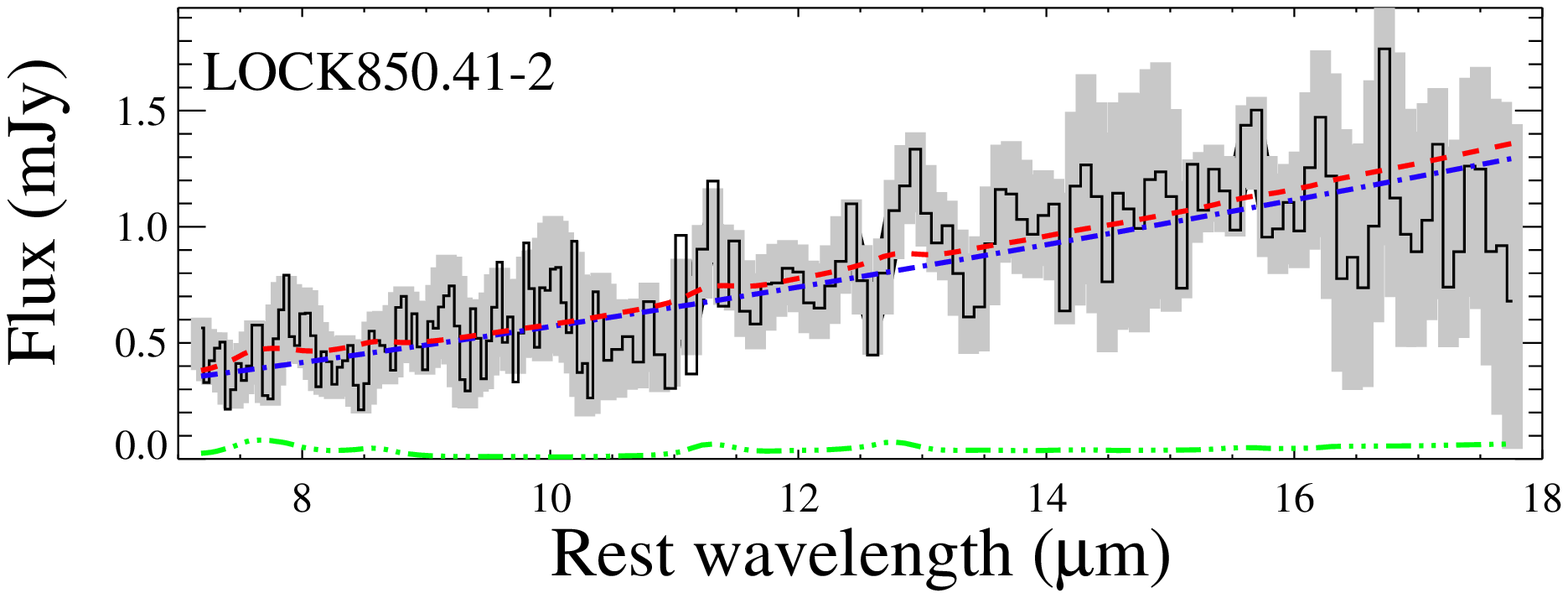}
\caption{The IRS spectrum of LOCK850.41-2: a continuum-dominated source in the mid-infrared ($<90$\% continuum/AGN fraction) with no visible PAH features from which a redshift could be obtained.  Instead we assume a source redshift of $z=0.974$, as indicated by the optical spectrum of the X-ray counterpart of the SMG from \citet{Lehmann01}.  The solid black histogram is the raw (unsmoothed) IRS spectrum, while the shaded region represents the associated $1\sigma$ noise from the sky background (see \S~\ref{sec:dr}).  The red dashed line is the best-fitting model composed of an M82 PAH template (green dot-dot-dot-dashed line) and a power-law component with extinction applied (blue dot-dashed line).  See \S~\ref{sec:analysis2}. \label{fig:lock41}}
\end{figure}

LOCK850.41 was first discovered in the 8\,mJy Survey (\citealt{Scott02}; LE850.08) and by the \citet{Greve04} 1200\,$\mu$m MAMBO survey (LH1200.014), and was reconfirmed by SHADES with an 850\,$\mu$m flux of 3.9$^{+0.9}_{-1.0}$\,mJy (deboosted) and a submm position of RA=10:51:59.86, Dec=57:24:23.6 \citep{Coppin06}. \citet{Ivison05,Ivison07} and \citet{Chapman05} located a statistically robust double counterpart in the radio for this system, as well as two coincidental 24\,$\mu$m counterparts, and obtained a convincing redshift of $z=0.689$ for the brightest, but furthest, of the counterparts.  A spectrum for the other radio counterpart was also obtained, containing broadly similar spectral features but at much lower S/N and so a redshift could not be unambiguously determined.  The redshift of the $z=0.689$ counterpart (LOCK850.41-1) was confirmed by \citetalias{KMD09} using the \textit{Spitzer}-IRS.  To complete the picture we obtained an IRS spectrum of LOCK850.41-2. This source originally followed our IRAC color selection criteria, although with a proper deblending of the IRAC photometry, it was found not to make our color selection cut after all, and so we have excluded this source from the main results and discussion of the paper.  In Fig.~\ref{fig:lock41}, the source appears to be completely continuum-dominated, with no strong discernable PAH features from which to determine a redshift.  We find a formal `minimum' $\chi^{2}$ best-fit at $z=3.1$ (see \S~\ref{sec:analysis1}) if the weak feature seen at $\sim11$\,$\mu$m is assumed to be the 6.2\,$\mu$m PAH feature, although we note that this solution is not unique as all the reduced $\chi^{2}$ values are $<1$ for the $0.5<z<4.5$ range that was considered.  This counterpart possesses a hard X-ray identification $\sim2.9''$ away ($S_{\mathrm{X\,(0.5-2\,keV)}}=17.5\times10^{-16}$\,erg\,cm$^{-2}$\,s$^{-1}$; $L_{\mathrm{X\,(2-10\,keV)}}=18.3\times10^{43}$\,erg\,s$^{-1}$; see Fig.~\ref{fig:stamps}) which is spectroscopically classified as a Type 1 AGN in the optical, which could explain the lack of visible strong PAH features in the IRS spectrum at $z=0.974$ \citep{Lehmann01}.  \citet{Mainieri02} find no evidence for X-ray absorption in this source, but do find a surprisingly flat X-ray spectral slope of $\Gamma\simeq0.7$, compared with a typical $\Gamma\simeq1.9$ for Type 1 AGN.  Despite the potential redshift ambiguity for LOCK850.41-2, the source is continuum-dominated in the mid-infrared, regardless of the redshift.  

\begin{thebibliography}{}
\bibitem[Alexander et al.(2001)]{Alexander01}Alexander D.M., Brandt W.N., Hornschemeier A.E., Garmire G.P., Schneider D.P., Bauer F.E., Griffiths R.E., 2001, \aj, 122, 2156 
\bibitem[Alexander et al.(2005a)]{Alexander05nat}Alexander D.M., Smail I., Bauer F.E., Chapman S.C., Blain A.W., Brandt W.N., Ivison R.J., 2005a, Nat., 434, 738
\bibitem[Alexander et al.(2005b)]{Alexander05}Alexander D.M., Bauer F.E., Chapman S.C., Smail I., Blain A.W., Brandt W.N., Ivison R.J., 2005b, ApJ, 632, 736
\bibitem[Alexander et al.(2008a)]{Alexander08a}Alexander D.M., et al., 2008a, AJ, 135, 1968
\bibitem[Alexander et al.(2008b)]{Alexander08}Alexander D.M., et al., 2008b, ApJ, 687, 835\bibitem[Archibald et al.(2002)]{Archibald02} Archibald, et al., 2002, MNRAS, 336, 353
\bibitem[Aretxaga et al.(2007)]{Aretxaga07}Aretxaga I. et al., 2007, MNRAS, 379, 1571
\bibitem[Armus et al.(2007)]{Armus07}Armus L. et al., 2007, ApJ, 656, 148
\bibitem[Ashby et al.(2006)]{Ashby06}Ashby M.L.N., et al., 2006, \apj, 644, 778
\bibitem[Austermann et al.(2010)]{Austermann09} Austermann J.E., et al., 2010, MNRAS, 401, 160
\bibitem[Blain et al.(2004)]{Blain04} Blain A.W., et al., 2004, ApJ, 611, 725
\bibitem[Borys et al.(2005)]{Borys05} Borys C., Smail I., Chapman S.C., Blain A.W., Alexander D.M., Ivison R.J., 2005, ApJ, 635, 853
\bibitem[Brandl et al.(2006)]{Brandl06}Brandl B.~R. et al., 2006, \apj, 653, 1129
\bibitem[Brunner et al.(2008)]{Brunner08}Brunner H., Cappelluti N., Hasinger G., Barcons X., Fabian A.C., Mainieri V., Szokoly G., 2008, A\&A, 479, 283
\bibitem[Chapin et al.(2009)]{Chapin09}Chapin E.L., et al., 2009, MNRAS, 398, 1793
\bibitem[Chapman et al.(2003)]{Chapman03}Chapman S.C., Blain A.W., Ivison R.J., Smail I.R., 2003, Nat., 422, 695
\bibitem[Chapman et al.(2005)]{Chapman05}Chapman S.C., Blain A.W., Smail I., Ivison R.J., 2005, ApJ, 622, 772
\bibitem[Chary \& Elbaz(2001)]{Chary01} Chary R., Elbaz D., 2001, ApJ, 556, 56
\bibitem[Clements et al.(2008)]{Clements08} Clements D.L. et al., 2008, MNRAS, 387, 247
\bibitem[Condon et al.(1992)]{Condon92}Condon J.J., 1992, ARA\&A, 30, 575
\bibitem[Coppin et al.(2006)]{Coppin06}Coppin K., et al., 2006, MNRAS, 372, 1621
\bibitem[Coppin et al.(2008a)]{Coppin08a}Coppin K., et al., 2008a, MNRAS, 384,1597
\bibitem[Coppin et al.(2008b)]{Coppin08b}Coppin K., et al., 2008b, MNRAS, 389, 45
\bibitem[Coppin et al.(2009)]{Coppin09}Coppin K., et al., 2009, MNRAS, 395, 1905
\bibitem[Dasyra et al.(2009)]{Dasyra09}Dasyra K.M., et al., 2009, ApJ, 701, 1123
\bibitem[Devlin et al.(2009)]{Devlin09}Devlin M.J., et al., 2009, Nat., 458, 737
\bibitem[Dey et al.(2008)]{Dey08}Dey A., et al., 2008, ApJ, 677, 943
\bibitem[Draine(2003)]{Draine03} Draine, B.~T. 2003, ARA\&A, 41, 241 
\bibitem[Draine \& Li(2007)]{Draine07} Draine, B.~T., \& Li, A.\ 2007, \apj, 657, 810 
\bibitem[Dye et al.(2008)]{Dye08}Dye S. et al., 2008, MNRAS, 386, 1107
\bibitem[Dye et al.(2009)]{Dye09}Dye S. et al., 2009, ApJ, 703, 285
\bibitem[Eales et al.(2009)]{Eales09}Eales S., et al., 2009, ApJ, 707, 1779
\bibitem[Farrah et al.(2008)]{Farrah08}Farrah D., et al., 2008, ApJ, 677, 957	 
\bibitem[Farrah et al.(2009b)]{Farrah09b}Farrah D., et al., 2009b, ApJ, 700, 395
\bibitem[Fazio et al.(2004)]{Fazio04}Fazio G.G., et al., 2004, ApJS, 154, 10
\bibitem[F\"{o}rster Schreiber et al.(2003)]{Forster03}F\"{o}rster Schreiber N.M., et al., 2003, A\&A, 399, 833
\bibitem[Frayer et al.(1998)]{Frayer98}Frayer D. T., Ivison R.J., Scoville N.Z., Yun M., Evans A.S., Smail I., Blain A. W., Kneib J.-P., 1998, ApJ, 506, L7
\bibitem[Furusawa et al.(2008)]{Furusawa08}Furusawa H. et al., 2008, ApJS, 176, 1
\bibitem[Genzel et al.(1998)]{Genzel98}Genzel R., et al., 1998, \apj, 498, 579
\bibitem[Greve et al.(2004)]{Greve04} Greve T.R. et al., 2004, MNRAS, 354, 779
\bibitem[Greve et al.(2005)]{Greve05} Greve T.R. et al., 2005, MNRAS, 359, 1165
\bibitem[Hern\'{a}n-Caballero et al.(2009)]{HC09} Hern\'{a}n-Caballero A., et al., 2009, MNRAS, 395, 1695
\bibitem[Hainline et al.(2009)]{Hainline09} Hainline L.J., Blain A.W., Smail I., Frayer D.T., Chapman S.C., Ivison R.J., Alexander D.M., 2009, ApJ, 699, 1610
\bibitem[Holland et al.(1999)]{Holland99} Holland W.S. et al., 1999, MNRAS, 303, 659
\bibitem[Houck et al.(2004)]{Houck04} Houck, J.~R., et al., 2004, \apjs, 154, 18 
\bibitem[Ibar et al.(2009)]{Ibar09}Ibar E., Ivison R.J., Biggs A.D., Lal D.V., Best P.N., Green D.A., 2009, MNRAS, 397, 281
\bibitem[Ibar et al.(2010)]{Ibar09b}Ibar E., Ivison R.J., Best P.N., Coppin K., Pope A., Smail I., Dunlop J.S., 2010, MNRAS, 401, L53
\bibitem[Ivison et al.(2004)]{Ivison04}Ivison R.J., et al., 2004, ApJS, 154, 124
\bibitem[Ivison et al.(2005)]{Ivison05}Ivison R.J., et al., 2005, MNRAS, 364, 1025
\bibitem[Ivison et al.(2007)]{Ivison07}Ivison R.J., et al., 2007, MNRAS, 380, 199
\bibitem[Ivison et al.(2008)]{Ivison08}Ivison R.J., et al., 2008, MNRAS, 390, 1117
\bibitem[Kennicutt(1998)]{Kennicutt98}Kennicutt R.~C., 1998, ARA\&A, 36, 189 
\bibitem[Kov\'{a}cs et al.(2006)]{Kovacs06}Kov\'{a}cs A., Chapman S.C., Dowell C.D., Blain A.W., Ivison R.J., Smail I., Phillips T.G., 2006, ApJ, 650, 592
\bibitem[Laird et al.(2010)]{Laird09}Laird E.S., Nandra K., Pope A., Scott D., 2010, MNRAS, 401, 2763
\bibitem[Laurent et al.(2000)]{Laurent00}Laurent O., et al., 2000, A\&A, 359, 887
\bibitem[Lawrence et al.(2007)]{Lawrence07}Lawrence A. et al., 2007, MNRAS, 379, 1599
\bibitem[Lehmann et al.(2001)]{Lehmann01} Lehmann I., et al., 2001, A\&A, 371, 833
\bibitem[Lilly et al.(1999)]{Lilly99}Lilly S.J., et al., 1999, \apj, 518, 641
\bibitem[Londsdale et al.(2004)]{Lonsdale04}Lonsdale C.J., et al., 2004, ApJS, 154, 54
\bibitem[Lutz et al.(2000)]{Lutz00}Lutz D., et al., 2000, \apj, 536, 697
\bibitem[Lutz et al.(2001)]{Lutz01}Lutz D., et al., 2001, A\&A, 378, 70
\bibitem[Lutz et al.(2004)]{Lutz04}Lutz D., Maiolino R., Spoon H.W.W., Moorwood A.F.M., 2004, A\&A, 418, 465 
\bibitem[Lutz et al.(2005)]{Lutz05}Lutz D., Valiante E., Sturm E., Genzel R., Tacconi L.J., Lehnert M.D., Sternberg A., Baker A.J., 2005, ApJ, 625, L83
\bibitem[Lutz et al.(2008)]{Lutz08}Lutz D. et al., 2008, ApJ, 684, 853
\bibitem[Mainieri et al.(2002)]{Mainieri02}Mainieri V., Bergeron J., Hasinger G., Lehmann I., Rosati P., Schmidt M., Szokoly G., Della Ceca R., 2002, A\&A, 393, 425
\bibitem[Men\'{e}ndez-Delmestre et al.(2007)]{MD07} Men\'{e}ndez-Delmestre K., et al., 2007, ApJ, 655, L65
\bibitem[Men\'{e}ndez-Delmestre et al.(2009)]{KMD09} Men\'{e}ndez-Delmestre K., et al., 2009, ApJ, 699, 667
\bibitem[Mortier et al.(2005)]{Mortier05} Mortier A., et al., 2005, MNRAS, 363, 563
\bibitem[Murphy et al.(2009)]{Murphy09}Murphy E.J., Chary R.-R., Alexander D.M., Dickinson M., Magnelli B., Morrison G., Pope A., Teplitz H.I., 2009, ApJ, 698, 1380
\bibitem[Peng et al.(2006)]{Peng06} Peng C.Y., et al., 2006, \apj, 649, 616
\bibitem[Pope et al.(2006)]{Pope06} Pope A., et al., 2006, MNRAS, 370, 1185
\bibitem[Pope et al.(2008a)]{Pope08} Pope A., et al., 2008a, \apj, 675, 1171
\bibitem[Pope et al.(2008b)]{Pope08b} Pope A., et al., 2008b, \apj, 689, 127
\bibitem[Rieke et al.(2004)]{Rieke04}Rieke G.H., et al., 2004, ApJS, 154, 25  
\bibitem[Rigopoulou et al.(1999)]{Rigopoulou99}Rigopoulou D., Spoon H.~W.~W., Genzel R., Lutz D., Moorwood A.~F.~M., Tran Q.~D., 1999, AJ, 118, 2625 
\bibitem[Sajina et al.(2007)]{Sajina07}Sajina A., Yan L., Armus L., Choi P., Fadda D., Helou G., Spoon H., 2007, ApJ, 664, 713
\bibitem[Salpeter(1955)]{Salpeter55}Salpeter E.E., 1955, \apj, 121, 161 
\bibitem[Sanders et al.(1988)]{Sanders88}Sanders D.B., Soifer B.T., Elias J.H., Neugebauer G., Matthews K., 1988, ApJ, 328, L35
\bibitem[Scott et al.(2002)]{Scott02}Scott S.E., et al., 2002, MNRAS, 331, 817
\bibitem[Siebenmorgen \& Kr\"{u}gel(2007)]{SK07}Siebenmorgen R., Kr\"{u}gel E., 2007, A\&A, 461, 445
\bibitem[Simpson et al.(2004)]{Simpson04}Simpson C., Dunlop J.S., Eales S.A., Ivison R.J., Scott S.E., Lilly S.J., Webb T.M.A., 2004, MNRAS, 353, 179
\bibitem[Skrutskie et al.(2006)]{Skrutskie06} Skrutskie M.F. et al.\ 2006, AJ, 131, 1163
\bibitem[Spergel et al.(2003)]{Spergel03}Spergel D.N. et al., 2003, ApJS, 148, 175
\bibitem[Springel, Di Matteo \& Hernquist(2005)]{Springel05}Springel V., Di\,Matteo T., Hernquist L., 2005, MNRAS, 361, 776
\bibitem[Stevens et al.(2005)]{Stevens05}Stevens J.A., Page M.J., Ivison R.J., Carrera F.J., Mittaz J.P.D., Smail I., McHardy I.M., 2005, MNRAS, 360, 610
\bibitem[Swinbank et al.(2004)]{Swinbank04} Swinbank, A. M., Smail, I., Chapman, S. C., Blain, A. W., Ivison, R. J., Keel, W. C., 2004, ApJ, 617, 64
\bibitem[Swinbank et al.(2006)]{Swinbank06} Swinbank, A. M., Chapman, S. C., Smail, I., Lindner, C., Borys, C., Blain, A. W., Ivison R. J., Lewis, G. F., 2006, MNRAS, 371, 465
\bibitem[Tacconi et al.(2006)]{Tacconi06}Tacconi L.J. et al., 2006, ApJ, 640, 228
\bibitem[Tacconi et al.(2008)]{Tacconi08}Tacconi L.J. et al., 2008, ApJ, 680, 246
\bibitem[Takata et al.(2006)]{Takata06}Takata T., Sekiguchi K., Smail I., Chapman S.C., Geach J.E., Swinbank A.M., Blain A., Ivison R.J., 2006, ApJ, 651, 713
\bibitem[Teplitz et al.(2007)]{Teplitz07} Teplitz, H.~I., et al., 2007, \apj, 659, 941 
\bibitem[Tran et al.(2001)]{Tran01}Tran, Q.~D., et al., 2001, \apj, 552, 527 
\bibitem[Ueda et al.(2008)]{Ueda08}Ueda Y., et al., 2008, ApJS, 179, 124
\bibitem[Valiante et al.(2007)]{Valiante07} Valiante E., et al., 2007, ApJ, 660, 1060 
\bibitem[Wall, Pope \& Scott(2008)]{Wall08}Wall J., Pope A., Scott D., 2008, MNRAS, 383, 435
\bibitem[Watabe et al.(2009)]{Watabe09}Watabe Y., Risaliti G., Salvati M., Nardini E., Sani E., Marconi A., 2009, MNRAS, 396, L1
\bibitem[Weedman et al.(2006)]{Weedman06} Weedman D.W., et al., 2006, ApJ, 653, 101
\bibitem[Wilson et al.(2008)]{Wilson08} Wilson G.W., et al., 2008, MNRAS, 386, 807
\bibitem[Yan et al.(2007)]{Yan07}Yan, L., et al., 2007, \apj, 658, 778
\bibitem[Yun et al.(2008)]{Yun08}Yun M.S., et al., 2008, MNRAS, 389, 333
\end{thebibliography}
\end{document}